\documentclass{aa}  

\usepackage{caption}
\usepackage{graphicx}
\usepackage{ulem}
\usepackage{txfonts}
\usepackage{color}
\definecolor{linkcolor}{rgb}{0,0,.7}
\definecolor{oldText}{rgb}{0.5,0,0}
\definecolor{newText}{rgb}{0,.5,0}

\usepackage[colorlinks=true]{hyperref}
\hypersetup{
  colorlinks   = true,
  urlcolor     = linkcolor,
  linkcolor    = linkcolor, 
  citecolor   = linkcolor 
}

\renewcommand{\H}{\text{H}}
\renewcommand{\O}{\text{O}}
\renewcommand{\doi}[1]{\textcolor{linkcolor}{\textbf{doi:}} \href{http://doi.org/#1}{#1}}

\newcommand{\scalinglaw}[1]{\left(\frac{#1}{#1_\oplus}\right)}
\newcommand{\CO}{\text{CO}_2}

\newcommand{\C}{\text{C}}

\newcommand{\PCO}{$P_{\text{CO}_2}$}

\newcommand{\der}[2]{\frac{d{#1}}{d{#2}}}
\newcommand{\e}[1]{\cdot 10^{#1}}
\newcommand{\parder}[2]{\frac{\partial{#1}}{\partial{#2}}}

\newcommand{\unit}[2]{\,\text{#1}^{#2}}

\begin{document} 

   \title{The role of planetary interior in the long-term evolution of atmospheric CO$_2$ on Earth-like exoplanets}

   \author{M. Oosterloo
          \inst{1}
          \and
          D. Höning
          \inst{2,3}
          \and
          I. E. E. Kamp
          \inst{1}
          \and
          F. F. S. van der Tak
          \inst{1,4}
          }

   \institute{Kapteyn Astronomical Institute, University of Groningen, Landleven 12, 9747 AD Groningen, The Netherlands
                  \and
Origins Center, Nijenborgh 7, 9747 AG Groningen, The Netherlands
               \and
               Department of Earth Sciences, Vrije Universiteit Amsterdam, De Boelelaan 1085,
1081 HV Amsterdam, The Netherlands
               \and
               SRON Netherlands Institute for Space Research, Landleven 12, 9747 AD Groningen, The Netherlands}

   \date{Received: 10-13-2020; accepted: 03-05-2021}

 
  \abstract
   {The long-term carbonate silicate cycle plays an important role in the evolution of Earth's climate and, therefore, may also be an important mechanism in the evolution of the climates of Earth-like exoplanets. However, given the large diversity in the possible interiors for Earth-like exoplanets, the ensuing evolution of the atmospheric CO$_2$ pressure may be widely different.}
   {We assess the role of the thermal evolution of the planetary interior on the long-term carbon cycle of Earth-like exoplanets. In particular, we investigate the effects of radiogenic mantle heating, core size, and planetary mass on the atmospheric partial CO$_2$ pressure, and the ability of a long-term carbon cycle driven by plate tectonics to control the atmospheric CO$_2$ pressure.}
   {We developed a box-model which connects carbon cycling to parametrized mantle convection. Processes considered in the carbon cycle are temperature-dependent continental weathering, seafloor weathering, subduction, and degassing through ridge and arc volcanism. The carbon cycle was coupled to the thermal evolution via the plate speed, which was parametrized in terms of the global Rayleigh number.}
   {We find decreasing atmospheric CO$_2$ pressure with time, up to an order of magnitude over the entire main sequence lifetime of a solar-type star. High abundances of radioactive isotopes allow for more efficient mantle degassing, resulting in higher CO$_2$ pressures. Within the spread of abundances found in solar-type stars, atmospheric CO$_2$ pressures at 4.5 Gyr were found to vary from $14$ Pa to $134$ Pa. We find a decreasing Rayleigh number and plate speed toward planets with larger core mass fractions $f_c$, which leads to reduced degassing and lower atmospheric CO$_2$ pressure. In particular for $f_c\gtrsim 0.8$, a rapid decrease of these quantities is found. Variations in planet mass have more moderate effects. However, more massive planets may favor the development of more CO$_2$ rich atmospheres due to hotter interiors.}
   {The dependence of plate tectonics on mantle cooling has a significant effect on the long-term evolution of the atmospheric CO$_2$ pressure. Carbon cycling mediated by plate tectonics is efficient in regulating planetary climates for a wide range of mantle radioactive isotope abundances, planet masses and core sizes. More efficient carbon cycling on planets with a high mantle abundance of thorium or uranium highlights the importance of mapping the abundances of these elements in host stars of potentially habitable exoplanets. Inefficient carbon recycling on planets with a large core mass fraction ($\gtrsim$ 0.8) emphasizes the importance of precise mass-radius measurements of Earth-sized exoplanets.}

    \keywords{planets and satellites: interiors -- planets and satellites: atmospheres -- planets and satellites: tectonics -- planets and satellites: terrestrial planets}

    \maketitle

    \section{Introduction}
    In the past two decades, astronomy has seen the emergence of the field of exoplanets, starting with the discovery of the first exoplanet around a solar-type star \citep{Mayor1995}. Since the discovery of this hot Jupiter, the number of discovered exoplanets has increased to over 4200, including planets which are Earth-like both in terms of mass and in terms of effective temperature \citep{Schneider2011, vanHoolst2019}. A key question with respect to these Earth-like exoplanets is whether these planets are habitable and whether they are able to remain habitable over timescales long enough for life to develop and evolve \citep[e.g.,][]{Kasting2003}.\\
\indent
Notable examples of potentially habitable exoplanets are Proxima Centauri b \citep{Anglada2016}, Trappist-1 e, f \& g \citep{Gillon2017}, and Teegarden's Star c \citep{Zechmeister2019}. However, these planets reside in tight orbits around red dwarfs, which confronts these exoplanets with environments widely different from Earth. Examples are the frequent stellar flares associated with red dwarfs \citep{Vida2017, Gunther2020} and the tidal interaction which likely leaves these planets tidally locked to their host star. These effects can compromise the habitability of exoplanets around M-type stars.\\
\indent
Rocky exoplanets are also detected in the habitable zones of solar-type\footnote{With "solar-type" or "Sun-like" we refer to main-sequence stars of spectral class G.} stars. An example is Kepler-452b, an exoplanet of $R\approx 1.6$ R$_\oplus$ which could be rocky in composition \citep{Jenkins2015}. Kepler-452b orbits around a solar-type star which is $\sim6\,\unit{Gyr}{}$ old, which is about 1.5 Gyr older than the Sun. As a consequence, Kepler-452b receives $\sim10\%$ more flux from its host star than Earth, which could result in the orbit of Kepler-452b to lie near the inner edge of the habitable zone inferred from conservative estimates \citep{Kopparapu2013, Jenkins2015}. This could mean Kepler-452b has entered a so-called moist-greenhouse state where most of the planetary water budget is lost to space \citep{Kasting1993, Abbott2012, Kopparapu2013}. However, as Sun-like stars such as Kepler-452 gradually increase in luminosity over time, Kepler-452b must have been cooler in the past, which could have allowed for the existence of surface liquid water for most of its lifetime \citep{Jenkins2015}. Altogether, the evolution of Sun-like stars can induce significant increases in the surface temperatures of orbiting planets, which may in turn compromise the long-term habitability of these planets.\\
\indent
A negative feedback mechanism which could offset the effects of brightening host stars is the long-term carbon cycle, which can act as a planetary thermostat \citep{Walker1981, Kasting1993}. This mechanism can control the evolution of the global climate over geological timescales via the recycling of CO$_2$ between the atmosphere and ocean, crust and mantle. The climate is regulated via temperature-dependent silicate weathering, which removes CO$_2$ from the atmosphere and provides a negative feedback mechanism for the surface temperature. If surface temperatures become lower, the weathering rate is suppressed, allowing CO$_2$ to build up in the atmosphere over time via volcanic outgassing, increasing surface temperatures. Conversely, high surface temperatures give rise to an increased weathering rate, removing more atmospheric CO$_2$ and therefore lowering surface temperatures. The carbon removed from the atmosphere is deposited on the seafloor as carbonate rocks. Subduction of the seafloor allows for the transport of carbon toward Earth's mantle, or to degas back into the atmosphere via arc volcanism. Altogether, the long-term carbon cycle is thought to have allowed Earth to maintain temperatures on its surface favorable for the evolution of life, despite the significant increase in solar luminosity since Earth's formation \citep{Sagan1972, Kasting1993}.\\
\indent 
Due to the importance of the carbon cycle for Earth's long-term habitability, the efficiency at which such a feedback mechanism operates is also relevant to consider for the habitability of Earth-like exoplanets. Here, efficiency refers to the possibility of carbon to cycle between the mantle and surface, and the timescale on which this process controls and regulates the atmospheric CO$_2$ pressure. Observational constraints on the characteristics of these exoplanets remain limited to mass-radius measurements. This makes the planetary characterization required to answer these questions challenging. However, these measurements do reveal a large diversity in the mass and bulk composition of Earth-like exoplanets. For example, Kepler-452b has been shown to likely have an interior that is composed of a larger faction of rock than Earth's \citep{Jenkins2015}. These differences in interior composition affect the thermal and chemical structure of the interior, and the composition and amount of volatile material outgassed into the atmosphere \citep{Noack2014b, Dorn2018, Ortenzi2020, Spaargaren2020}. The thermal budget of a planet is affected by the rate of radiogenic heating, which predominantly occurs via the decay of the long-lived radioactive refractory elements thorium and uranium \citep{Schubert2001}. A first-order estimate for the abundances of the radioactive isotopes in the mantles of exoplanets are the abundances found in the atmosphere of the host star \citep{Unterborn2015}. Up to now, the implications of this diversity in interior structure for the efficiency of long-term carbon cycling and feasibility as a climate-regulation mechanism remain poorly understood.\\
\indent 
Crucial for a carbon cycle similar to Earth's is the presence of plate tectonics, which facilitates the transport of carbon into the mantle via subduction. In addition, plate tectonics replenishes atmospheric CO$_2$ via volcanic degassing at mid-oceanic ridges and subduction zones \citep{Kasting2003}. The plate speed and prevalence of plate tectonics on Earth-like exoplanets, however, remain poorly understood \citep{Berovici2015}. While some modeling studies suggest plate tectonics is generally likely on Earth-like exoplanets \citep{Valencia2007, Foley2012, Tackley2013}, other studies disagree \citep{Oneill2007b, Kite2009, Stamenkovic2012, Noack2014}. The uncertainty in the requirements for plate tectonics to emerge is typically circumvented in planetary evolution models by imposing the presence \citep[e.g.,][]{Foley2015} or absence \citep[e.g.,][]{Tosi2017, Foley2018, Foley2019} of plate tectonics.\\
\indent
Even when assuming the presence of plate tectonics, the diversity in the interiors of Earth-like exoplanets leaves a large parameter space to explore. This originates from the fact that plate tectonics is the surface manifestation of convection in Earth's mantle \citep[e.g.,][]{Schubert2001}, which in turn depends on the thermal and chemical structure of the mantle \citep[e.g.,][]{Spaargaren2020}. In addition, the manner in which plate tectonics couples to these interior properties remains poorly understood. Even for Earth, the history of plate tectonics is not fully characterized \citep{Palin2020}. Usually studies considering planets in the plate tectonics regime explore the influence of a single variable on the long-term carbon cycle, such as mantle temperature \citep{Sleep2001}, surface temperature \citep{Foley2015} or decarbonization fraction in subduction zones \citep{Hoening2019b}.\\
\indent
In this work we assessed the effects of the planetary interior on long-term carbon cycling facilitated by plate tectonics. More specifically, we investigated the role of radioactive isotope abundance, core size and mass on the evolution of the atmospheric partial CO$_2$ pressure. In addition, we assessed the range of these parameters for which plate tectonics provides an efficient mechanism for carbon recycling, while we also derived an indication of the minimum age of a planet required for its atmospheric CO$_2$ pressure to be completely regulated by the long-term carbon cycle. For these purposes, we developed a two-component model which connects a parametrized mantle convection model to a box model describing the evolution of the long-term carbon cycle. Plate tectonics here acts as the interface between thermal evolution and carbon cycling, where we parametrized the mean plate speed in terms of the Rayleigh number \citep{Schubert2001, Turcotte2014}.\\
\indent 
We elaborate on our model setup in Sect. \ref{sec:2}, where we derive a parametrization for the plate speed to connect mantle convection with carbon cycling. Subsequently, we present our key results for different radioactive isotope abundance, core size and planet mass in Sect. \ref{sec:3}, whose validity is discussed in Sect. \ref{sec:4}. In Sect. \ref{sec:5} we discuss the implications of our results for the long-term evolution of atmospheric CO$_2$ pressure on Earth-like exoplanets. Finally, we summarize our key conclusions in Sect. \ref{sec:6}.

    \section{Model setup}
    \label{sec:2}

In this section, we present condensed discussions of the thermal evolution and carbon cycling model in Sects. \ref{sec:2.1} and \ref{sec:2.2}, while a full technical discussion is provided in Appendix \ref{sec:AA} and \ref{sec:AB}, respectively. Subsequently, we motivate and derive a scaling law for the plate speed as the main link between interior and carbon cycle in Sect. \ref{sec:2.3}, while Sect. \ref{sec:2.4} presents the scaling laws required to subsequently extrapolate our model to planets of different core size and mass. 

\subsection{Thermal evolution model}
\label{sec:2.1}
\begin{figure}
    \centering
    \includegraphics[width=.49\textwidth]{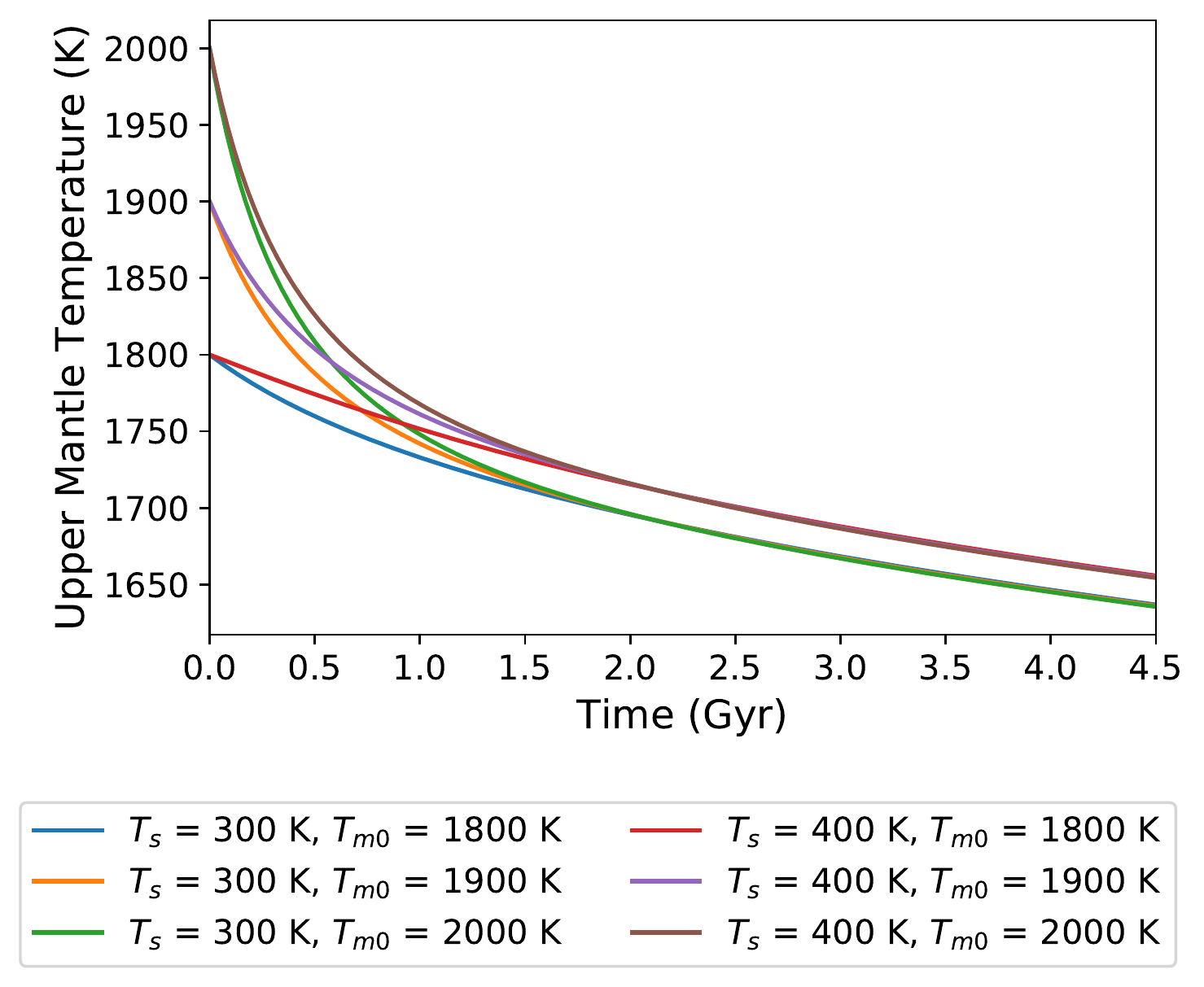}
    \caption{Evolution of the upper mantle temperature $T_\text{m}$ for different surface temperatures $T_\text{s}$ and initial upper mantle temperatures $T_\text{m0}$.}
    \label{fig:21MantleTemperature}
\end{figure}

The key aim of our thermal evolution model is to provide a first-order estimate of the evolution of the vigor of mantle convection over time, which both directly and indirectly depends on mantle temperature \citep[e.g.][]{Schubert2001}. Therefore, usage of a parametrized mantle convection model is appropriate. Our model is similar to the models considered in \cite{Driscoll2014} and \cite{Hoening2016}. A full discussion of our model is presented in Appendix \ref{sec:AA}.\\
\indent
We restrict ourselves to a two-component planetary composition; an iron core surrounded by a magnesium-silicate mantle, as these two planetary components are thought to be the main constituents of terrestrial planets \citep{Seager2007, Zeng2013}. The mantle is assumed to consist of a single, isoviscous, convective layer. The convective mantle is heated externally from below by the core and internally by the decay of radioactive isotopes. Cooling occurs via heat loss though a conductive thermal boundary layer at the surface.\\
\indent The mantle viscosity plays an important role, as it controls the long-term convective cooling of the mantle. This is demonstrated in Fig. \ref{fig:21MantleTemperature}, where the evolution of the upper mantle temperature $T_\text{m}$ is shown for a set of different initial mantle temperatures ($T_\text{m0}=1800, 1900, 2000$ K), while the surface temperature is fixed at a constant value ($T_\text{s}=300, 400$ K). Additional model parameter values are presented in Table \ref{tab:TETable}. The initial mantle temperature only plays a role in the first 1-2 Gyr, while later, the evolution of the mantle temperature is governed by the viscosity, independent of initial mantle temperature. On the other hand, the constant surface temperature $T_\text{s}$ continues to play a role throughout the entire thermal history due to its effect on the surface heat flux $q_\text{u}$. Higher surface temperatures result in a hotter mantle. 

\subsection{Carbon cycle model}
\label{sec:2.2}
As a next step, we quantify the effects of the thermal evolution on the atmospheric CO$_2$ pressure with a box model of the carbon cycle. We here follow a modeling approach similar to, for example, \cite{Tajika1992}, \cite{Sleep2001}, and \cite{Foley2015}. A more comprehensive discussion of the model itself is presented in Appendix \ref{sec:AB}.\\
\indent
For the carbon cycle we consider the partition of a total carbon budget over carbon reservoirs for the atmosphere, ocean, oceanic crust, and mantle. The evolution of the carbon content of these reservoirs is found by calculating the carbon added and removed from the reservoirs at each timestep, while the total carbon budget remains conserved. We assume equilibrium of carbon between the atmosphere and ocean reservoir to be instantaneous, as the timescale for equilibrium between these reservoirs is short ($\sim 10^3$ yr) compared to the other reservoirs ($\gtrsim 10^6$~yr) \citep{Sleep2001}. Atmospheric and oceanic carbon is removed via both continental and seafloor weathering, and stored on the seafloor in the form of carbonate rocks. Subduction leads to carbon sequestration in the mantle, while a fixed fraction is degassed in the atmosphere via arc volcanism. In addition, mantle carbon is degassed directly in the atmosphere at mid oceanic ridges. Subduction and ridge degassing both depend on the plate speed. Therefore, the rate at which carbon cycles through the interior is governed by the plate speed and the efficiency of the above outlined weathering processes.

\subsection{Plate speed parametrization}
\label{sec:2.3}
\begin{figure}
    \begin{center}
    \includegraphics[width=.49\textwidth]{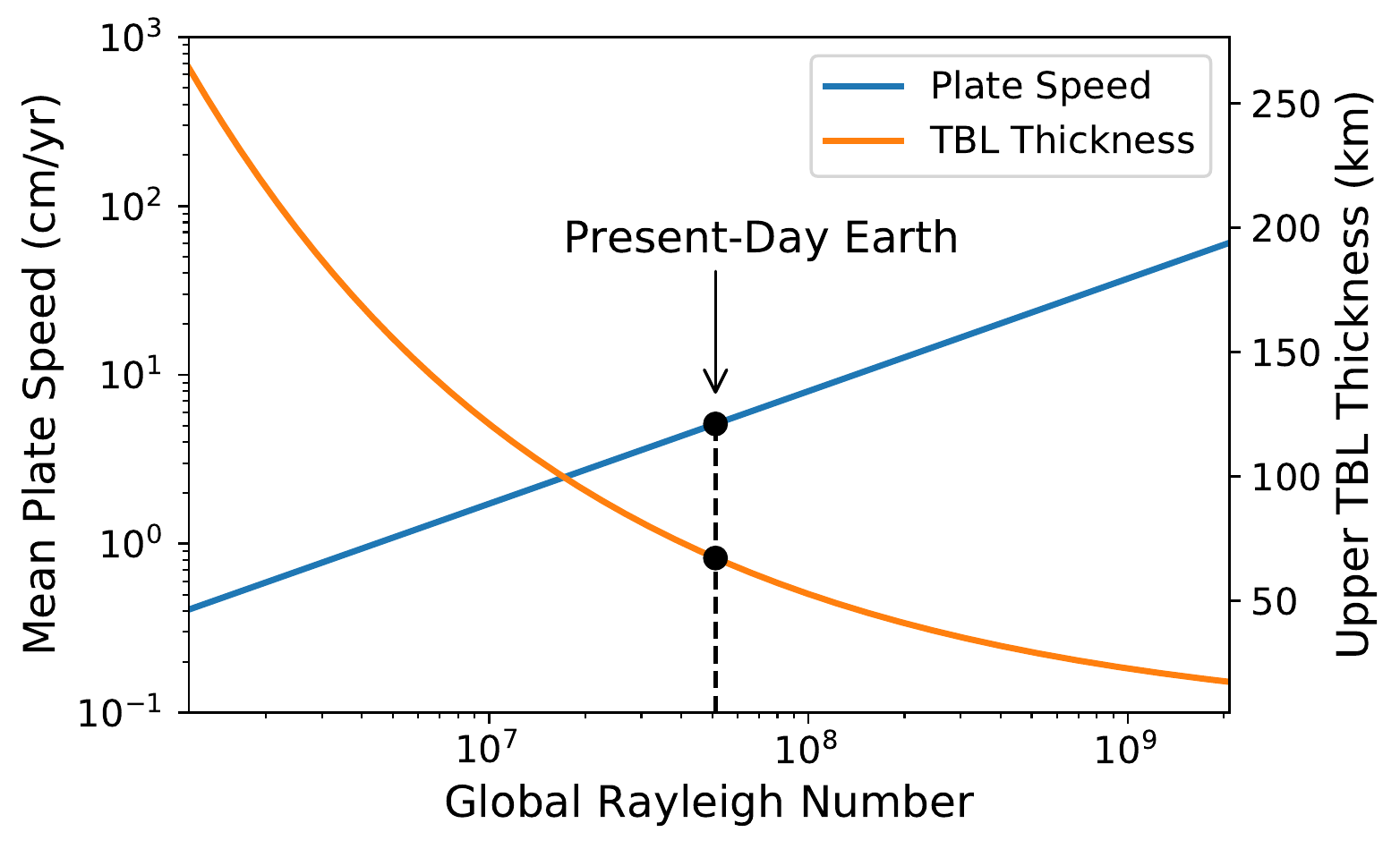}
    \caption{Plate speed scaling law given in Eq. \ref{eq:plateSpeedScalingLaw} as a function of the Rayleigh number Ra, along with the thickness of the upper thermal boundary layer (TBL). Values for present-day Earth are indicated for comparison.}
    \label{fig:23PlateSpeedDemo}
    \end{center}
\end{figure}
\begin{figure*}
    \centering
    \includegraphics[width=\textwidth]{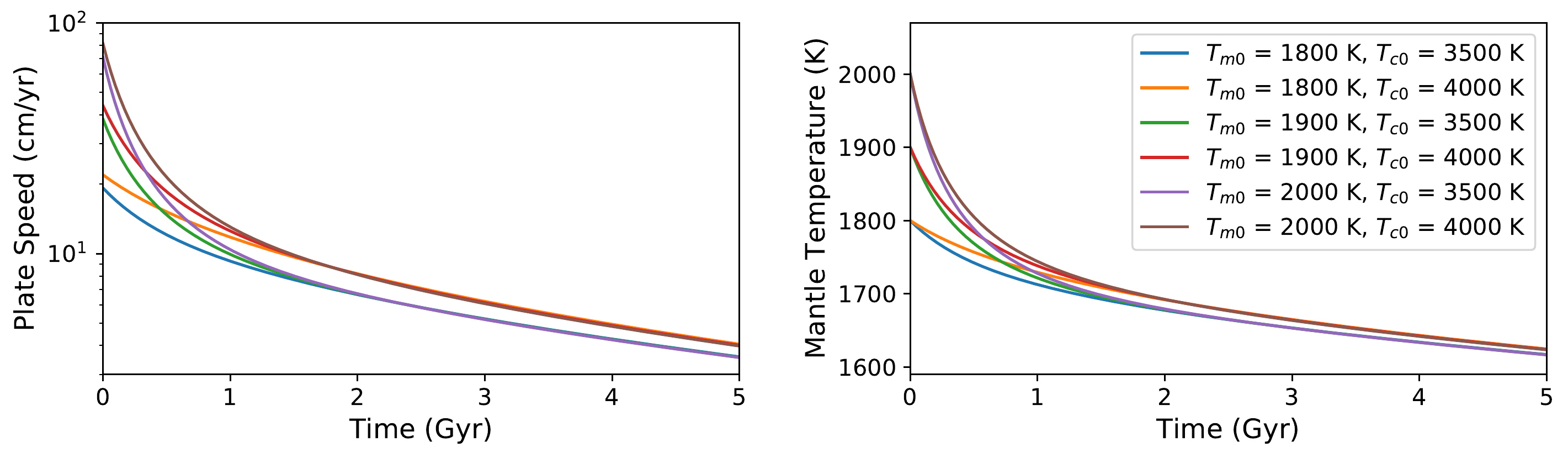}
    \caption{Evolution of plate speed $v_\text{p}$ (left) and mantle temperature $T_\text{m}$ (right) for different initial mantle and core temperatures.}
    \label{fig:23PlateSpeedAnalysis}
\end{figure*}

\begin{figure}
    \centering
    \includegraphics[width=.49\textwidth]{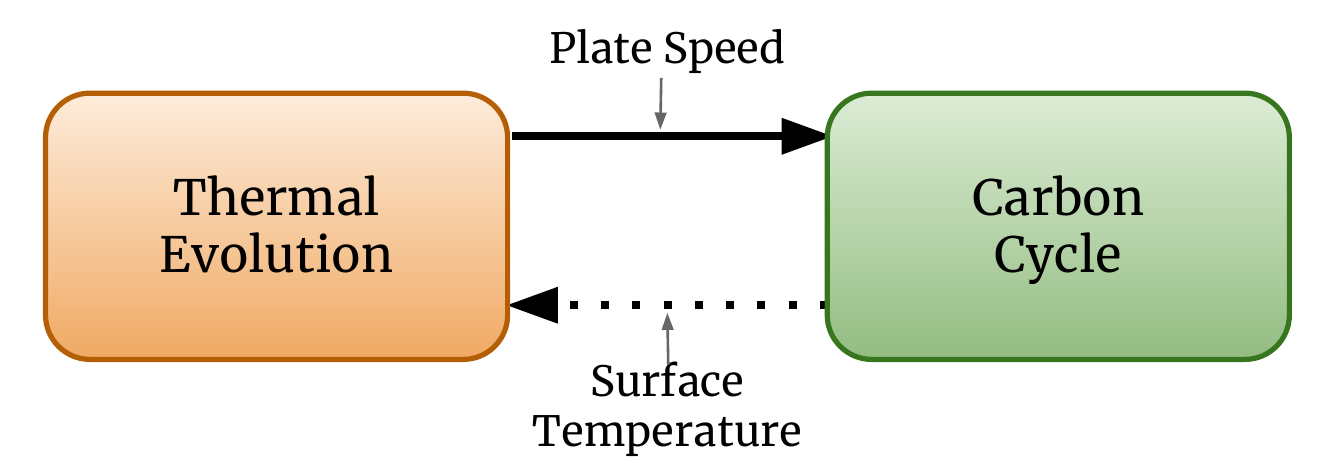}
    \caption{A visual summary of the model considered in this study.}
    \label{fig:23ModelOverview}
\end{figure}

In this study, we consider the plate speed $v_\text{p}$ as the main coupling variable between the carbon cycle and mantle convection models. It affects the rate at which carbon is transported toward the mantle via subduction, and from the mantle via degassing at mid-oceanic ridges. We motivate and derive here a parametrization for the plate speed in terms of the mantle global Rayleigh number and mantle thickness. The behavior of other couplings and their potential effects on the results of this study are discussed in Sect. \ref{sec:4.3}.\\
\indent
In the boundary layer model used to develop the thermal evolution model (Appendix \ref{sec:AA}), one can express the mean horizontal flow speed $u_\text{0}$ of mantle fluid in the thermal boundary layer near the planetary surface in terms of the mantle Rayleigh number \citep{Schubert2001, Turcotte2014}:
\begin{align}
\label{eq:fluidSpeed}
    u_\text{0}\,\propto\,\frac{1}{D}\text{Ra}^{2\beta}.
\end{align}
Here, $D$ denotes the mantle thickness, while $\beta$ denotes the Nusselt-Rayleigh coupling exponent, and describes the relation between mantle convection and the mantle cooling. Mantle convection is quantified by the Rayleigh number Ra, while mantle cooling is described by the Nusselt number Nu, the ratio between total heat flow and conductive heat flow \citep{Turcotte2014}. $\beta$ connects the Rayleigh and Nusselt number via
\begin{align}
\label{eq:NuRa}
    \text{Nu}\,\propto\,\text{Ra}^\beta.
\end{align}
Theoretical considerations of Rayleigh-Bénard convection with an isoviscous mantle fluid result in $\beta=\frac{1}{3}$ for Earth's mantle \citep{Turcotte2014}. However, the value of $\beta$ may be lower depending on mantle rheology (\citeauthor{Schubert2001} \citeyear{Schubert2001}; \citeauthor{Oneil2020} \citeyear{Oneil2020} and references therein). How the precise value of $\beta$ changes in the mantles of Earth-like exoplanets therefore remains unknown. Throughout this study, a value of $\beta=\frac{1}{3}$ is used, while we explore the effects of lower values of $\beta$ on our model results in Sect. \ref{sec:4.2}.\\
\indent 
We approximate the Rayleigh number with the global Rayleigh number of the mantle
\begin{align}
\label{eq:Rastart}
    \text{Ra}=\frac{g\rho_\text{m}^2c_\text{m}\alpha\left[(T_\text{c}-T_\text{s})-(T_\text{b}-T_\text{m})\right]D^3}{k\eta}.
\end{align}
Here, $g$ denotes the mean mantle gravity, $\rho_\text{m}$ the mean mantle density, $c_\text{m}$ the mantle heat capacity, $\alpha$ the thermal expansivity, and $k$ the thermal conductivity. In addition, $T_\text{c}$ denotes the temperature at the core-mantle boundary (see also Fig. \ref{fig:AATemperatureProfileOverview}), $T_\text{b}$ the temperature above the thermal boundary layer between the mantle and core, $T_\text{m}$ the mantle temperature below the upper thermal boundary layer, and $T_\text{s}$ the surface temperature. $\eta=\eta(T_\text{m})$ is the mantle dynamic viscosity, whose dependence on mantle temperature $T_\text{m}$ is described by the Arrhenius law given in Eq. \ref{eq:viscosity}.\\
\indent 
As the upper thermal boundary layer directly corresponds to the lithosphere, we assume that the plate motion is coupled to the motion of the mantle fluid, which implies that the plate speed is linearly coupled to the fluid speed:
\begin{align}
\label{eq:propTo}
    v_\text{p}\,\propto\,u_\text{0}.
\end{align}
Although this relation is generally applied to describe plate tectonics on present-day Earth \citep{Schubert2001}, the dynamics of the plates with respect to the mantle fluid in reality depends on the material properties of the lithosphere and mantle. In this context, \cite{Crowley2012} found Eq. \ref{eq:propTo} to be applicable whenever the material strength of the plates is low, such that the plate motion is fully controlled by the convective properties of the mantle. In contrast, if the plate material strength becomes high or the mantle viscosity low, other regimes exist for the plate speed behavior. In these regimes, the plate speed is less sensitive to the properties of the mantle. Instead of mantle properties, lithospheric material properties become an important or even dominant factor in controlling the plate dynamics. Therefore, Eq. 4 is valid only if the mantle and lithosphere properties are similar to the present-day Earth. We discuss the validity and implications of Eq. \ref{eq:propTo} more elaborately in Sect. \ref{sec:4.3}.\\ 
\indent
Using Eqs. \ref{eq:fluidSpeed} and \ref{eq:propTo}, we write the following scaling relation for the plate speed:
\begin{align}
\label{eq:plateSpeedScalingLaw}
    v_\text{p}=v_{\text{p}\oplus}\scalinglaw{D}^{-1}\scalinglaw{\text{Ra}}^{2\beta}.
\end{align}
Here $D_{\oplus}$ and Ra$_\oplus$ denote the present-day Earth mantle thickness and Rayleigh number, respectively, and are given in Table \ref{tab:TETable}. The value of Ra$_\oplus$ is inferred from a run of the uncoupled thermal evolution model with $T_\text{s}=T_{\text{s}\oplus}=285$ K, $T_\text{m0}=2000$ K, and $T_\text{c0}=3500$ K. The resulting behavior of the plate speed as a function of global Rayleigh number is presented in Fig. \ref{fig:23PlateSpeedDemo}.\\ 
\indent
It follows from Eqs. \ref{eq:Rastart} and \ref{eq:plateSpeedScalingLaw} that the plate speed has a strong dependence on mantle temperature $T_\text{m}$ through the mantle viscosity $\eta$, as given by Eq. \ref{eq:viscosity}. In addition, the plate speed also depends on core temperature $T_\text{c}$ via Eq. \ref{eq:Rastart}. Therefore it is to be expected that the initial mantle temperature $T_\text{m0}$ and core temperature $T_\text{c0}$ leave their mark on the evolution of the mean plate speed $v_\text{p}$. In order to assess the role of initial interior temperatures in more detail, we perform model test runs where the surface temperature is kept fixed at $T_\text{s}=285$ K, and show the evolution of the plate speed and mantle temperature for $T_\text{m0}=1600, 1800, 2000$ K and $T_\text{c0}=3500, 4000$ K in Fig. \ref{fig:23PlateSpeedAnalysis}. In the long term, all initial conditions result in a gradually declining plate speed at $t\gtrsim 1$ Gyr. Furthermore, the evolution of the plate speed up to $t\sim 2$ Gyr depends on both the initial mantle and core temperature, while their influence becomes significantly less pronounced at later times. Figure \ref{fig:23PlateSpeedAnalysis} also demonstrates that the evolution of the plate speed is primarily determined by $T_\text{m}$, both due to its effect on the temperature gradient in Eq. \ref{eq:Rastart} and the strong effect of $T_\text{m}$ on the mantle viscosity. On the other hand, the evolution of the core temperature $T_\text{c}$ affects the evolution of the plate speed both directly via the temperature gradient, and indirectly through the modified evolution of mantle temperature $T_\text{m}$.\\
\indent
In addition to the interior-carbon cycle coupling through the mean plate speed, our model contains another coupling between the CO$_2$ pressure and the thermal evolution of the interior via surface temperature $T_\text{s}$ (Fig. \ref{fig:23ModelOverview}). However, initial tests revealed that large ($\gtrsim 100$ K) differences in surface temperature must be maintained for prolonged periods of time ($\sim 1$ Gyr) to induce significant differences in the long-term evolution of the mantle temperature and plate speed. Therefore, we do not further discuss the role of this coupling in the rest of this work.

\subsection{Scaling mass and core size}
\label{sec:2.4}
\begin{figure*}
    \centering
    \includegraphics[width=\textwidth]{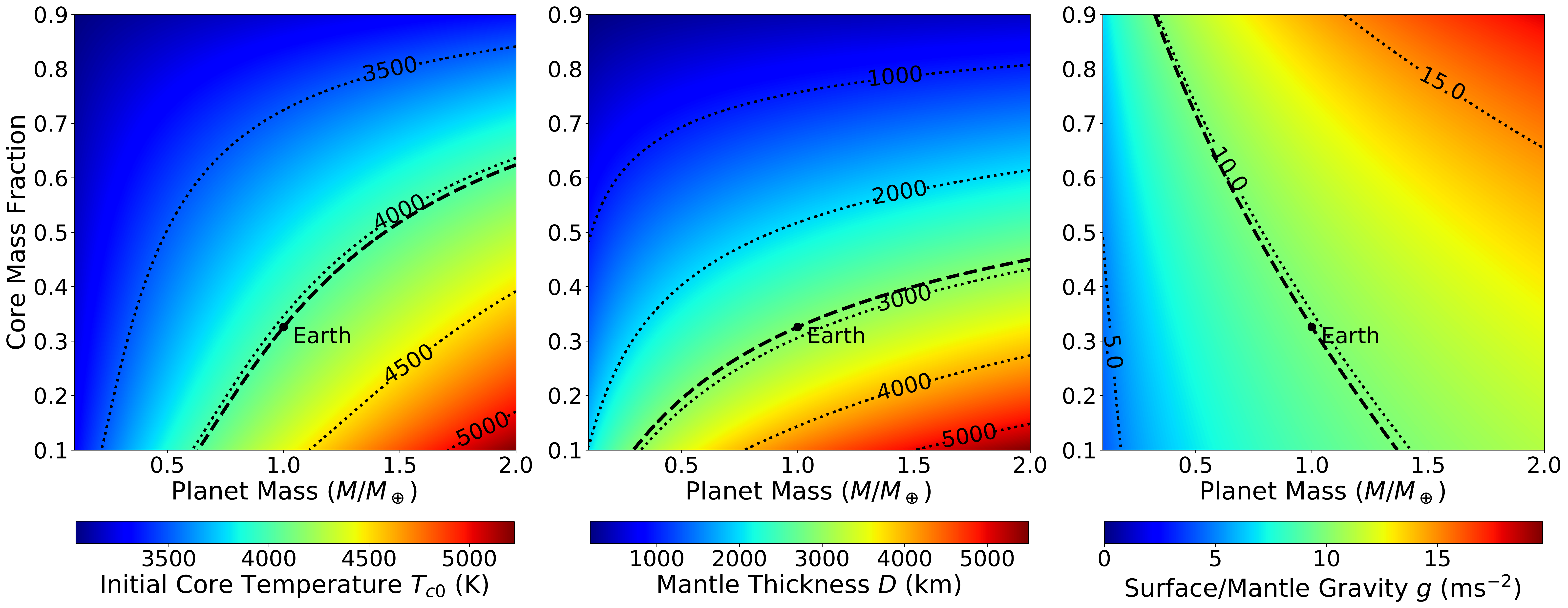}
    \caption{Initial core temperature $T_\text{c0}$ (left), mantle thickness $D$ (center), and surface gravity $g$ (right) as a function of core mass fraction $f_\text{c}$ and planet mass $M_p$.}
    \label{fig:24ConstantsDepiction}
\end{figure*}

The planet mass and core size have a large influence on the thermal structure and evolution of the mantle \citep{Valencia2006}. Therefore, we aim to assess how variations in these two planetary properties modify the long-term carbon cycle. For this purpose, we introduce parameter scalings in this section to extrapolate the model discussed so far toward planets with a core size and mass different from Earth.\\
\indent 
An important variable in this context is the mean planetary density $\bar{\rho}$, which is a function of planet composition \citep{Zapolsky1969}. We note that the density depends on the equation of state of the mantle and core material, and is thus a function of the temperature profile $T(r)$ and pressure profile $P(r)$ of the mantle and core \citep{Valencia2006, Unterborn2016}. For simplicity, however, we impose a mean mantle density $\rho_\text{m}$ and mean core density $\rho_\text{c}$ representative for the present-day Earth mantle and core. This simplification limits the parameter space that can be probed reliably. Even more restrictive is the isoviscous approximation, for the mantle, which forces us to consider only planets in the mass range from $0.1$ M$_\oplus$ (about the mass of Mars, \citeauthor{dePater2015} \citeyear{dePater2015}) to $2$ M$_\oplus$. We elaborate on this upper boundary of the planetary mass range in Sect. \ref{sec:4.1}. For the core mass fractions $f_\text{c}$, we consider values between 0.1 to 0.9; an interval which is motivated from the numerical instability of our model at large and small core mass fractions. We note however that pure silicate ($f_\text{c}=0$) or pure iron ($f_\text{c}=1$) exoplanets may exist \citep{Elkins-Tanton2008, vanHoolst2019}.\\
\indent 
The mean mantle density and core density are calculated via
\begin{align}
    \rho_\text{c}=\frac{f_{\text{c}\oplus}M_\oplus}{V_{\text{c}\oplus}},\qquad \rho_\text{m}=\frac{(1-f_{\text{c}\oplus})M_\oplus}{V_{\text{m}\oplus}}.
\end{align}
Here, $f_{c\oplus}\approx0.3259$ denotes Earth's core mass fraction \citep{Valencia2006} and $M_\oplus\approx 5.97\e{24}$ kg Earth's mass. Furthermore, $V_{m\oplus}$ and $V_{c\oplus}$ represent the volume of Earth's mantle and core, respectively, whose values along with the resulting densities are given in Table \ref{tab:TETable}.\\
\indent 
We approximate the mean mantle gravity $g$ with the planet's surface gravity, given by $g=GM_\text{p}/R_\text{m}^2$. In addition, we scale the total carbon budget $R_\text{tot}$ with mantle mass
\begin{align}
R_\text{tot}=\frac{(1-f_\text{c})}{(1-f_{c\oplus})}\frac{M_\text{p}}{M_\oplus}R_{\text{tot},\oplus}.
\end{align}
Here, we choose $R_{\text{tot},\oplus}=2.5\e{22}$ mol as the reference value for Earth \citep{Sleep2001}. Furthermore, we set the total length of mid-ocean ridges or subduction zones $L$ proportional to the planet radius $R_\text{m}$.\\
\indent
Another aspect to consider upon extrapolating is the initial mantle and core temperature throughout $M_\text{p}$-$f_\text{c}$ parameter space. Figure \ref{fig:23PlateSpeedAnalysis} shows that the choice of initial mantle temperature plays a large role in the early evolution of the plate speed. The initial temperature throughout the planetary interior is primarily set by an energy balance during accretion between the release of gravitational energy upon the formation and the radiative cooling of the outer surface of the planet \citep{vanHoolst2019}. Subsequently, differentiation provides additional interior heat. In our model, we assume an Earth-like lithosphere and core to be already present at $t=0$. Temperature gradients throughout the lithosphere have been found to be only a weak function of planet mass; $T_\text{m}-T_\text{s} \sim M_\text{p}^{0.02}$ \citep{Valencia2007}. Using this scaling relation, one finds the upper mantle temperature $T_\text{m}$ in a 10 $M_\oplus$ super Earth to differ less than 5\% with respect to a 1 $M_\oplus$ (otherwise identical) planet. Altogether, we choose an initial upper mantle temperature of $T_\text{m0}=2000$ K as a first order approximation to the initial mantle temperature, following for example \cite{Noack2014}. The latter also impose a temperature increase $\Delta T_\text{c}$ of roughly $1000$~K between the bottom mantle temperature $T_\text{b}$ and temperature at the core-mantle boundary $T_\text{c}$. We impose a similar initial temperature gradient through the lower thermal boundary layer as an initial condition. Altogether this allows us to write $T_\text{c0}$ in terms of $T_\text{m0}$, using that $T_\text{m}$ increases adiabatically to $T_\text{b}$ as given by Equation \ref{eq:TB}:
\begin{align}
\label{eq:initialCMB}
    T_\text{c0}&\approx T_\text{m0}\left(1+\frac{\alpha g(D-2\delta_\text{u})}{c_\text{m}}\right)+\Delta T_\text{c}.
\end{align}
With this expression, we find values for $T_\text{c0}$ that are comparable with the values for $T_\text{c0}$ used in \cite{Noack2014}. However, we note that the value of $\Delta T_\text{c}$ and also the temperature profile relating $T_\text{m}$ to $T_\text{b}$ in reality depend on the physical and chemical properties of the mantle. For example, the thermal structure depends on the local behavior of the viscosity, which is in reality not only a function of temperature, but also of mantle material composition and pressure \citep{Stamenkovic2011, Stamenkovic2012}. The high pressures thought to prevail in the lower mantles of massive super Earths could substantially increase the local viscosity. This would in turn impair the efficiency of convection as a heat transport mechanism, and could lead to super-adiabatic temperature profiles in the deeper mantle \citep{Tackley2013}. For simplicity, these effects are not considered in our thermal evolution model. Instead, we limit our study to planets where we expect these effects to be negligible to first order. Therefore, we consider planets up to a maximum mass $M_p\leq 2 M_\oplus$; a limit which we motivate more elaborately in Sect. \ref{sec:4.1}. We note that the dependence of $T_\text{c0}$ on planet mass and core size are contained implicitly in Eq. \ref{eq:initialCMB} via the mantle thickness $D\equiv R_\text{m}-R_\text{c}$ and mean mantle gravity $g$. In addition, $D$ and $g$ directly affect the Rayleigh number and plate speed through Eqs. \ref{eq:Rastart} and \ref{eq:plateSpeedScalingLaw}. Therefore, by varying planet mass and core mass fraction, we effectively vary the mantle thickness and mean mantle gravity in this study.\\ 
\indent
We explore the behavior of $T_\text{c0}$, $D$ and $g$ as a function of planet mass and core mass fraction in Fig. \ref{fig:24ConstantsDepiction}. The initial core temperature remains between 3000 and 4000 K for most of the parameter space considered ($0.1<f_\text{c}<0.9$; $0.1M_\oplus\leq M_\text{p} \leq 2M_\oplus$). An exception exists for more massive planets with a core fraction below 0.5, where $T_\text{c0}$ can exceed 5000 K with $f_\text{c}$ approaching 0.1. This behavior is a result of the dependence of $T_\text{c0}$ on mantle thickness. The variation of the mantle thickness, shown in the center panel of Fig. \ref{fig:24ConstantsDepiction}, displays a rapid increase in mantle thickness at small $f_\text{c}$. Though this effect applies to any planet due to the assumed mass-radius relation, the effect is more pronounced for more massive planets due to their larger size. Mantle gravity, which also affects $T_\text{c0}$, is larger for massive planets with large core mass fractions.

    \section{Results}
    \label{sec:3}
This section presents the findings of our model. As a first step, we assess how the distribution of carbon over the various reservoirs evolves over time under the influence of mantle cooling in Sect. \ref{sec:3.1}. Subsequently we explore the effects of different interior properties on the long term carbon cycle: mantle radioactive isotope abundance (Sect. \ref{sec:3.2}), core size (Sect. \ref{sec:3.3}) and planet mass (Sect. \ref{sec:3.4}). Thereafter, we investigate the feasibility of carbon cycling mediated by plate tectonics throughout this parameter space in Sect. \ref{sec:3.5}, where we introduce the equilibrium timescale as a diagnostic tool.

\subsection{Basic behavior}
\label{sec:3.1}
\begin{figure*}
    \centering
    \includegraphics[width=\textwidth]{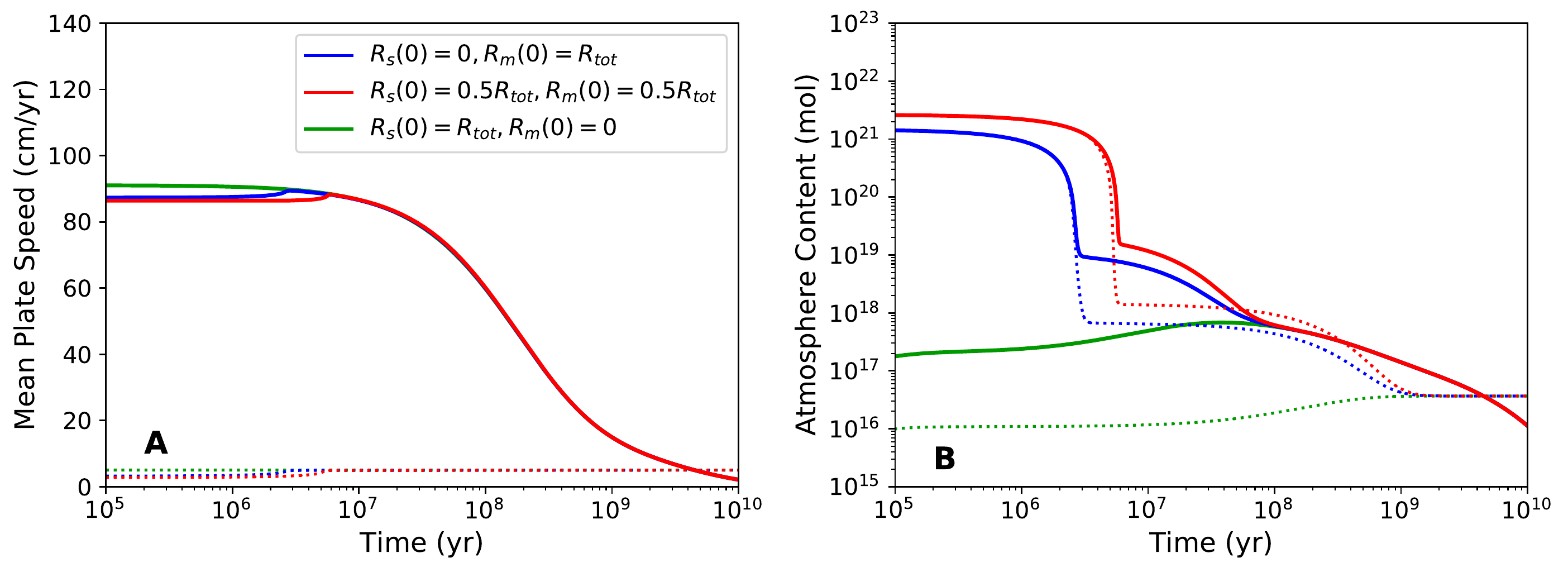}
\end{figure*}
\begin{figure*}
    \centering
    \includegraphics[width=\textwidth]{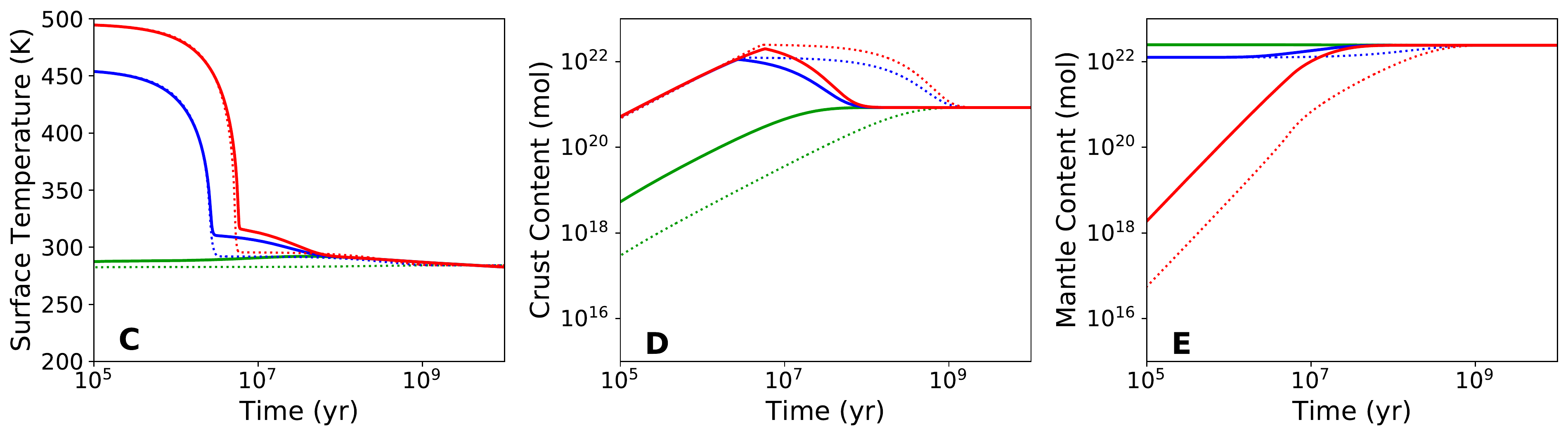}
    \caption{Evolution of plate speed $v_\text{p}$ (A), atmosphere reservoir content $R_\text{a}$ (B), surface temperature $T_\text{s}$ (C), crust reservoir content $R_\text{k}$ (D) and mantle reservoir content $R_\text{m}$ (E) for different initial distributions of the total carbon budget $R_\text{tot}$ over the various reservoirs. Different color depict model runs with different initial surface reservoirs $R_\text{s}(t=0)=R_\text{a}(t=0)+R_\text{o}(t=0)$.} Solid lines indicate results from this work, while dotted lines represent results from \cite{Foley2015}. 
    \label{fig:31PrimaryResultsComparison}
\end{figure*}
Figure \ref{fig:31PrimaryResultsComparison} shows the effects of mantle cooling on the long-term carbon cycling, where the evolution of the plate speed, surface temperature and carbon reservoirs are presented for three different initial distributions of carbon over the various reservoirs. In addition, we investigate differences of our results with respect to \cite{Foley2015}, where the plate speed is assumed to be a function of surface temperature. Test runs revealed that differences between the evolution obtained from the model of \cite{Foley2015} and a model where the plate speed is fixed at its present-day value of $5\,\unit{cm}{}\unit{yr}{-1}$ are negligible in most cases. Noticeable differences were only found when the surface temperature is very high, resulting in lower plate speeds in the model of \cite{Foley2015}. The total budget of carbon participating in the carbon cycle is kept constant at $R_\text{tot}=2.5\e{22}$~mol, while the initial conditions consist of a case where all carbon is initially in the surface reservoir $R_\text{s0}=R_\text{a0}+R_\text{o0}=R_\text{tot}$, in the mantle reservoir $R_\text{m0}=R_\text{tot}$, and an intermediate case where $R_\text{s0}=R_\text{m0}=\frac{1}{2}R_\text{tot}$. Furthermore, we assume $T_\text{m0}=2000$ K and $T_\text{c0}=3500$ K as initial conditions for the thermal evolution model. These initial conditions serve as our baseline initial conditions throughout the rest of this work, unless stated otherwise. The solar irradiation $S_\text{irr}(t)$ is kept fixed at its present-day value $S_\odot$ in order to isolate the effects of $S_\text{irr}=S_\text{irr}(t)$ and by the cooling interior on the CO$_2$ pressure and $T_\text{s}$. A discussion of the effects of $S_\text{irr}(t)$ is presented in Appendix \ref{sec:AC}. All model parameters are presented in Tables \ref{tab:TETable} and \ref{tab:CCTable}.\\
\indent
For the two cases with initially CO$_2$-rich atmospheres, the evolution of the atmospheric CO$_2$ pressure can be classified into three phases: Initially, weathering removes carbon from the atmosphere and it gradually accumulates on the seafloor (Fig.~\ref{fig:31PrimaryResultsComparison}B\&D). Despite the initially high surface temperatures (up to 500 K), liquid water, a requirement for weathering, is still present on the surface due to the higher surface pressure resulting from the thick, CO$_2$-rich atmosphere \citep{Foley2015}. However, due to the high surface temperature and atmospheric CO$_2$ pressure, continental weathering is initially supply-limited: the weathering rate is limited by the physical erosion rate, and not on the kinetics of the weathering reaction \citep[e.g.][and references therein]{Foley2015}. This means continental weathering initially proceeds at a rate $F_{\text{w}_\text{s}}$, constant with respect to surface temperature and atmospheric CO$_2$ pressure (see Appendix \ref{sec:AB}). After $10^6$ to $10^7$ yr, equilibrium between arc volcanism and weathering is established, while carbon continues to build up in the mantle reservoir (Fig. \ref{fig:31PrimaryResultsComparison}E) until equilibrium between both ridge degassing and arc volcanism and weathering is established after around $10^8$ yr. In this last step, the plate speed plays a pivotal role as it governs the rate at which carbon can be sequestered in the mantle. In the case where the carbon budget is fully in the mantle, the plate speed is also the limiting factor which determines the time required for equilibrium between the various reservoirs to be established as it controls degassing from the mantle reservoir.\\
\indent
We initially find the plate speed to be significantly larger than its present-day Earth value (Fig. \ref{fig:31PrimaryResultsComparison}A) ($\sim 90\,\unit{cm}{}\unit{yr}{-1}$ compared to $\sim 5\,\unit{cm}{}\unit{yr}{-1}$) due to the initially hot mantle in our model. This causes the time until equilibrium between the reservoirs is established to decrease by an order of magnitude (from $\sim10^9$ to $\sim10^8$ yr) with respect to the model of \cite{Foley2015}.

\subsection{Influence of radiogenic heating}
\label{sec:3.2}
\begin{figure*}
    \centering
    \includegraphics[width=\textwidth]{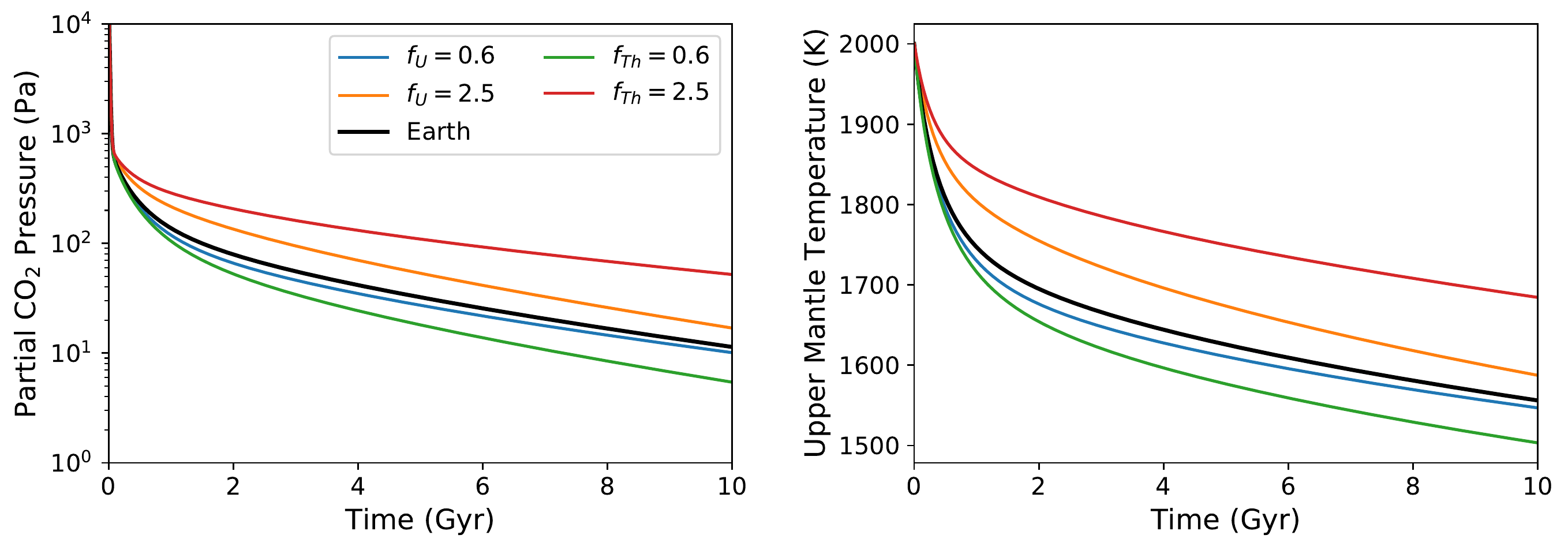}
   \caption{Evolution of partial CO$_2$ pressure \PCO (left) and upper mantle temperature $T_\text{m}$ (right) for different abundances of uranium ($^{235}$U and $^{238}$U) and thorium ($^{232}$Th). Abundances are expressed relative to Earth ($f_\text{U}=f_\text{Th}=1$).}
    \label{fig:32IsotopeComparison}
\end{figure*}

As a next step, we investigate the effects of radiogenic heating on carbon cycling. For this purpose we consider the long-term evolution of partial CO$_2$ pressure \PCO\, and mantle temperature $T_\text{m}$ in Fig. \ref{fig:32IsotopeComparison} for different abundances of uranium and thorium, which scale linearly with their specific heat productions (parameters $q_i$ in Table \ref{tab:TETable}). The initial distribution of carbon is set to $R_\text{s0}=R_\text{tot}$, although the effects of this initial condition become negligible within the first $10^8$ yr. Furthermore, all initial conditions and parameter values are kept fixed with respect to section \ref{sec:3.1}. For scaling the radiogenic heating provided by the decay of thorium, we use the constraints derived by \cite{Unterborn2015}, and scale the thorium abundance $f_\text{Th}$ between 0.6 and 2.5 Earth's thorium abundance. We also apply these limits for the total uranium abundance $f_\text{U}=f_\text{235U}+f_\text{238U}$ as the nucleosynthetic origin of thorium and uranium are correlated \citep{Goriely2001, Frebel2007, Unterborn2015}. \\
\begin{figure*}
    \centering
    \includegraphics[width=\textwidth]{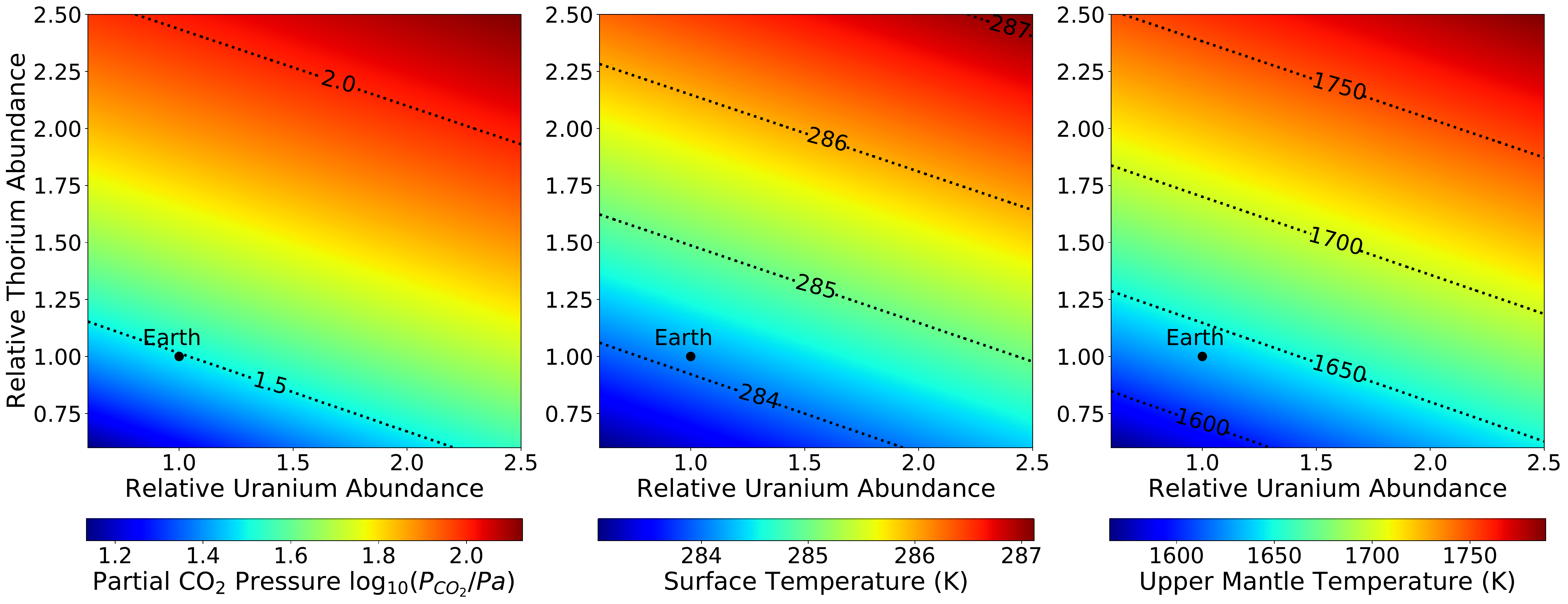}
    \caption{Present-day partial CO$_2$ pressure \PCO (left), surface temperature $T_\text{s}$ (center) and mantle temperature $T_\text{m}$ (right) as a function of thorium and uranium abundance relative to Earth.}
    \label{fig:32IsotopeFullSpace}
\end{figure*}
\indent
Figure \ref{fig:32IsotopeComparison} reveals that increased radioactive isotope abundances result in a higher mantle temperature and hence CO$_2$ pressure due to the effects of mantle temperature on the plate speed via the Rayleigh number. In addition, a high abundance of uranium results in a more rapid decline of \PCO\,over time. The mantle temperature appears to display a stronger decline for $f_\text{U}=2.5$, due to the larger contribution of uranium to mantle heating at earlier times. The evolution of \PCO\,follows this more rapid mantle cooling. In addition, differences between models with differing radioactive isotope variations tend to grow over time. Quantitatively, this amounts at $t=10$ Gyr to a temperature ranging from 1500 K to 1680 K for the coolest and hottest mantle, respectively. This range of mantle temperatures mentioned above translates into a partial CO$_2$ pressure ranging from 5 Pa to 45 Pa ($t=10$ Gyr); almost one order of magnitude. It also becomes clear that variations in the abundance of thorium have a larger effect on the thermal evolution and resulting atmospheric CO$_2$ pressure than uranium, which can be explained by thorium's long half-life and comparatively high abundance in Earth's crust. Enrichment or depletion of uranium in the mantle has a similar effect on the mantle temperatures and \PCO, albeit less pronounced, in particular at later times. This effect results from the shorter half-life of the uranium isotopes.\\
\indent
As a next step, we investigate the implications of a different isotopic abundance for the planetary climate. In Fig. \ref{fig:32IsotopeFullSpace}, we consider snapshots at $t=4.5$ Gyr of \PCO, $T_\text{s}$ and $T_\text{m}$ as a function of $f_\text{Th}$ and $f_\text{U}$. Both abundances are varied within the constraints provided by \cite{Unterborn2015}. For $f_\text{Th}=f_\text{U}=2.5$ we find a mantle temperature of about 1790 K, compared to 1560 K for the case where $f_\text{Th}=f_\text{U}=0.6$. This results in \PCO\,ranging from $\sim 14$ Pa to $\sim 134$ Pa. The average surface temperature ranges from $283$ K to $287$ K. Furthermore, Fig. \ref{fig:32IsotopeFullSpace} reveals that $T_\text{m}$ does not increase linearly as a function of radioactive isotope abundance. Instead, the mantle temperature (and hence \PCO\,and $T_\text{s}$) appears to increase more slowly if the amount of radiogenic heating is larger. Overall, a higher isotopic abundance results in a hotter mantle. This in turn enhances the vigor of convection in the mantle due to the decreasing viscosity, resulting in enhanced degassing. Therefore, one can expect to find more long-term accumulation of CO$_2$ in atmospheres of planets with high abundances of long-lived radioactive isotopes.
\begin{figure*}
    \centering
    \includegraphics[width=\textwidth]{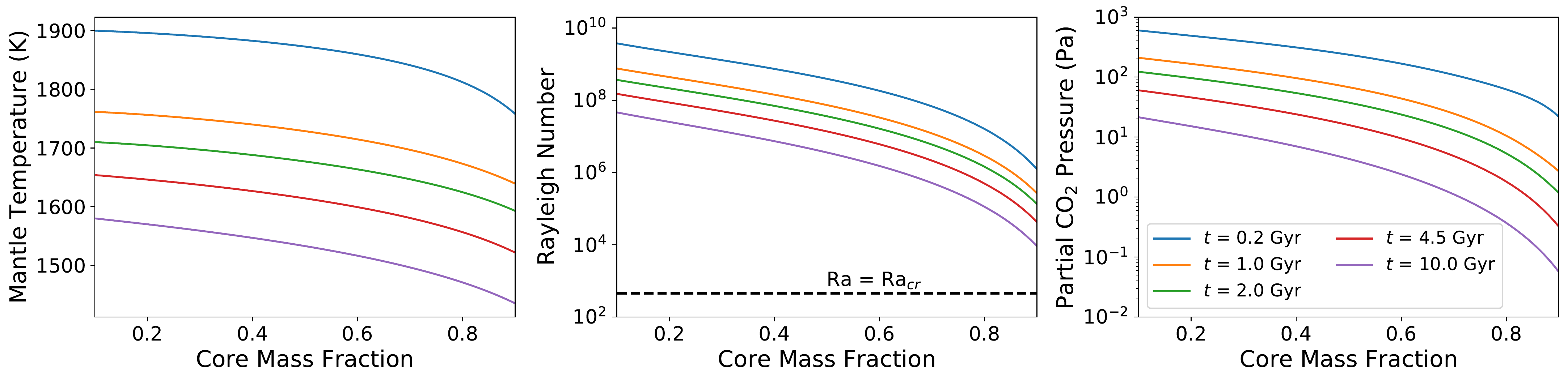}
    \caption{Snapshots over time of upper mantle temperature $T_\text{m}$ (left), the global Rayleigh number (center) and partial CO$_2$ pressure \PCO\, (right) as a function of core mass fraction $f_\text{c}$.}
    \label{fig:33CoreSizeEvolution}
\end{figure*}
\begin{figure*}
    \centering
    \includegraphics[width=\textwidth]{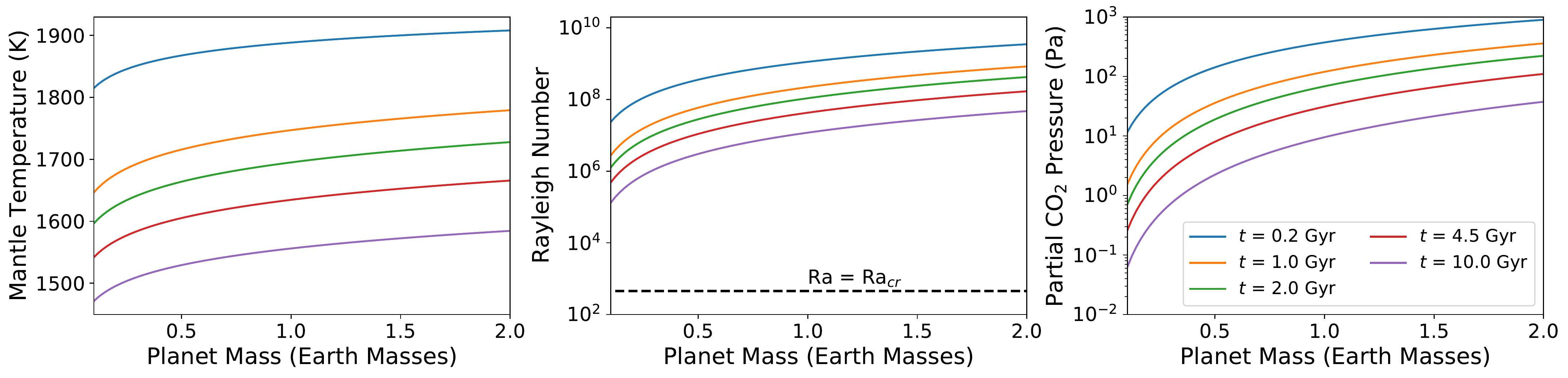}
    \caption{Snapshots over time of upper mantle temperature $T_\text{m}$ (left), the global Rayleigh number (center) and partial CO$_2$ pressure \PCO\, (right) as a function of planet mass $M_\text{p}$.}
    \label{fig:34MassEvolution}
\end{figure*}

\subsection{Influence of core size}
\label{sec:3.3}
In Fig. \ref{fig:33CoreSizeEvolution}, we explore the effects of the core size on carbon cycling by taking snapshots at $t=0.1,1,2,4.5$ and $10$ Gyr of the mantle temperature, Rayleigh number and atmospheric CO$_2$ pressure as a function of core mass fraction $f_\text{c}$. The core mass fraction is varied between 0.1 and 0.9. This parameter range is motivated from the numerical stability of our model. The former fraction corresponds to a mostly silicate planet which consists largely of a mantle. It has been argued that coreless planets may exist \citep{Elkins-Tanton2008}. On the other hand exoplanets with an inferred mean density consistent with pure iron have been found as well, which motivates our choice of the upper boundary \citep{vanHoolst2019}. All planets considered are assumed to have an initially CO$_2$ rich atmosphere (initially, $R_\text{s}=R_\text{a}+R_\text{o}=R_\text{tot}$), and an initial mantle temperature of 2000~K.\\
\indent
The mantle temperatures shown in Fig. \ref{fig:33CoreSizeEvolution} indicate that the mantles of planets with massive cores cool faster; a result from their larger surface to volume ratio. The faster cooling results in a rapidly increasing viscosity which reduces the Rayleigh number via Eq. \ref{eq:Rastart}. Additionally, a thinner mantle automatically results in a lower Rayleigh number due to the cubic dependence of the Rayleigh number on mantle thickness $D$. Altogether this results in a rapidly declining Rayleigh number for core fractions $f_\text{c}\gtrsim 0.8$; a trend followed by \PCO\, due to the dependence of the plate speed on the Rayleigh number. As the Rayleigh number scales with $D^3$, the plate speed scales as $v_\text{p}\sim D^{-1}\cdot(D^3)^{2/3}=D$ for $\beta=\frac{1}{3}$. Therefore, the plate speed also decreases sharply at large core fractions. As a result, the rates of seafloor weathering, ridge volcanism and subduction also become severely limited, and the mantle reservoir decouples from the surface reservoir. On the other hand, continental weathering continues to transport CO$_2$ from the atmosphere and ocean to the seafloor, resulting in a strongly decreasing atmospheric CO$_2$ pressure at large core fractions. We note this result is comparable to the conclusions of \cite{Noack2014b} for stagnant lid planets. In addition, the lower mantle temperature and more sluggish plate movement at large core fractions could imply that plate tectonics itself is also more difficult to initiate on those planets in the first place. Altogether a large planetary core can be detrimental for the feasibility of a long-term carbon cycle facilitated by plate tectonics, as it severely disrupts the exchange of carbon between the mantle and surface.

\subsection{Influence of planet mass}
\label{sec:3.4}
In Fig. \ref{fig:34MassEvolution} we consider snapshots of the mantle temperature, Rayleigh number and CO$_2$ pressure as a function of planet mass $M_\text{p}$ at different timesteps. We consider planets with a mass of $0.1$~M$_\oplus\leq M_\text{p}\leq 2$~M$_\oplus$; a parameter range which we discuss in more detail in Sect. \ref{sec:4.1}. \\
\indent We find the atmospheric CO$_2$ pressure to increase toward more massive planets due to the increasing mantle thickness. This results in the mantle of a $2$ M$_\oplus$ planet at $t=4.5$ Gyr to be $\sim30$ K hotter than that of a $1$ M$_\oplus$ planet. The lower viscosity results in a larger Rayleigh number, which promotes degassing via the plate speed. The resulting CO$_2$ pressure at $t=4.5$ Gyr ranges from $\sim 30$ Pa to $\sim 110$ Pa for a $1$ M$_\oplus$ and $2$ M$_\oplus$ planet, respectively.\\
\indent
Toward less massive planets, the mantle thickness decreases strongly, as demonstrated in Fig. \ref{fig:24ConstantsDepiction}. Therefore, the mantle is able to cool more efficiently. This is in particular true for planets with $M_\text{p}<0.5$ M$_\oplus$, below which we see an accelerated decrease in mantle temperature. A $0.1$ M$_\oplus$ planet has a mantle which is roughly $\sim 90$ K cooler than a $1$ M$_\oplus$ planet at $t=4.5$ Gyr, resulting in a partial CO$_2$ pressure of only $\sim 0.3$ Pa.\\
\indent 
For the overall efficiency of the carbon cycle, however, variations in planetary mass appear to have more moderate effects than the core size. This is in particular true for planets which are more massive than Earth. This is a result from the assumed mass-radius relationship $M_\text{p}\propto R_\text{m}^{1/3}$, which causes the mantle surface-volume ratio $A_\text{m}/V_\text{m}$ to decrease more slowly toward massive planets. However, we note this cooling behavior as a function of $M_\text{p}$ is a matter of debate, and may depend on the Nu-Ra coupling parameter $\beta$ (Eq. \ref{eq:NuRa}) \citep{Stevenson2003, Seales2020, Seales2021}. Altogether, we find that the effects of planetary mass on the thermal evolution and hence carbon cycle are more moderate toward more massive planets.

\subsection{Equilibrium timescales}
\label{sec:3.5}
\begin{figure}
    \centering
    \includegraphics[width=.4\textwidth]{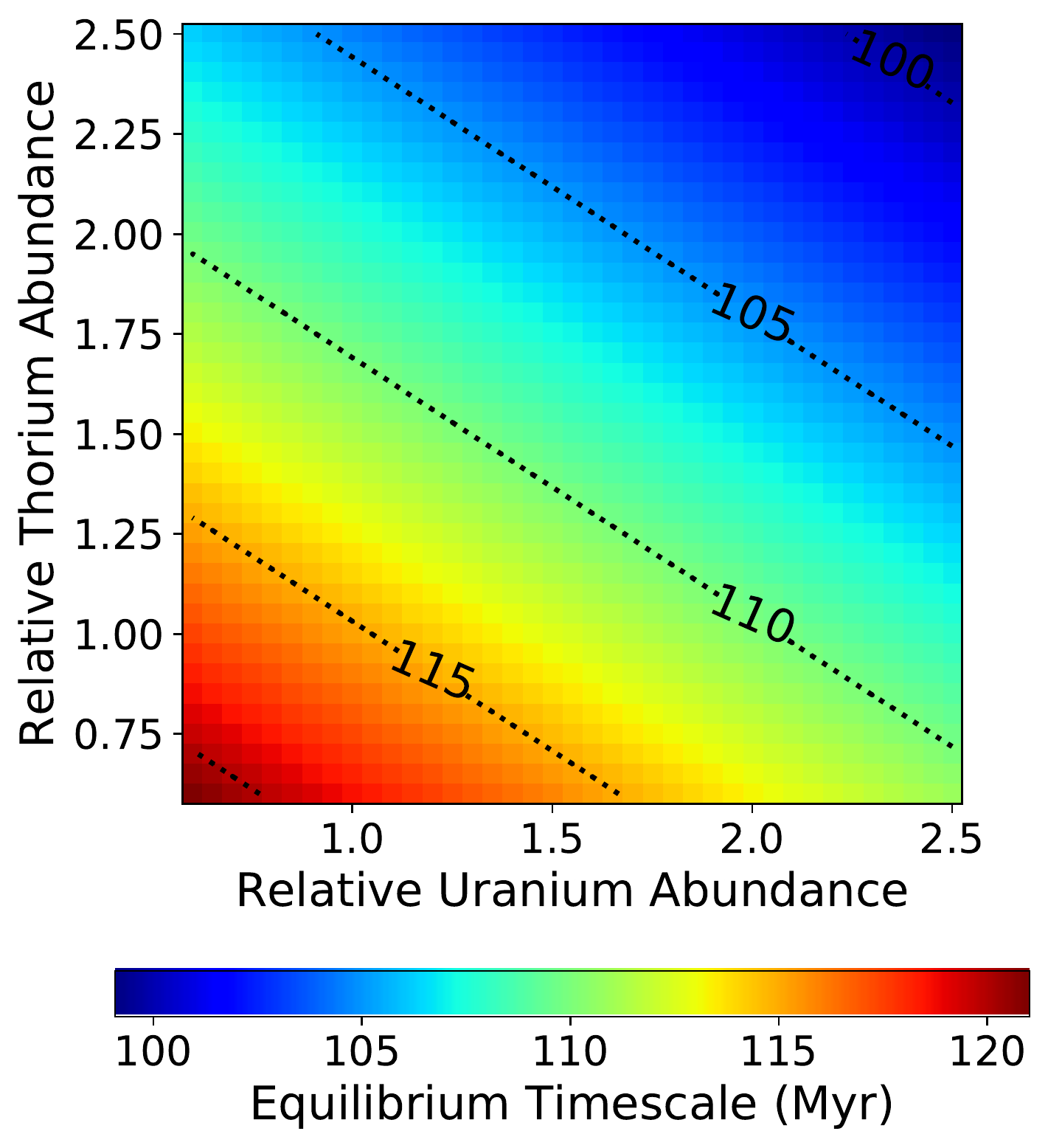}
    \caption{Equilibrium timescale $\tau_\text{CC}$ as a function of mantle abundance of uranium ($^{235}$U and $^{238}$U) and thorium.}
    \label{fig:35IsotopeTimescales}
\end{figure}
\begin{figure}
    \centering
    \includegraphics[width=.4\textwidth]{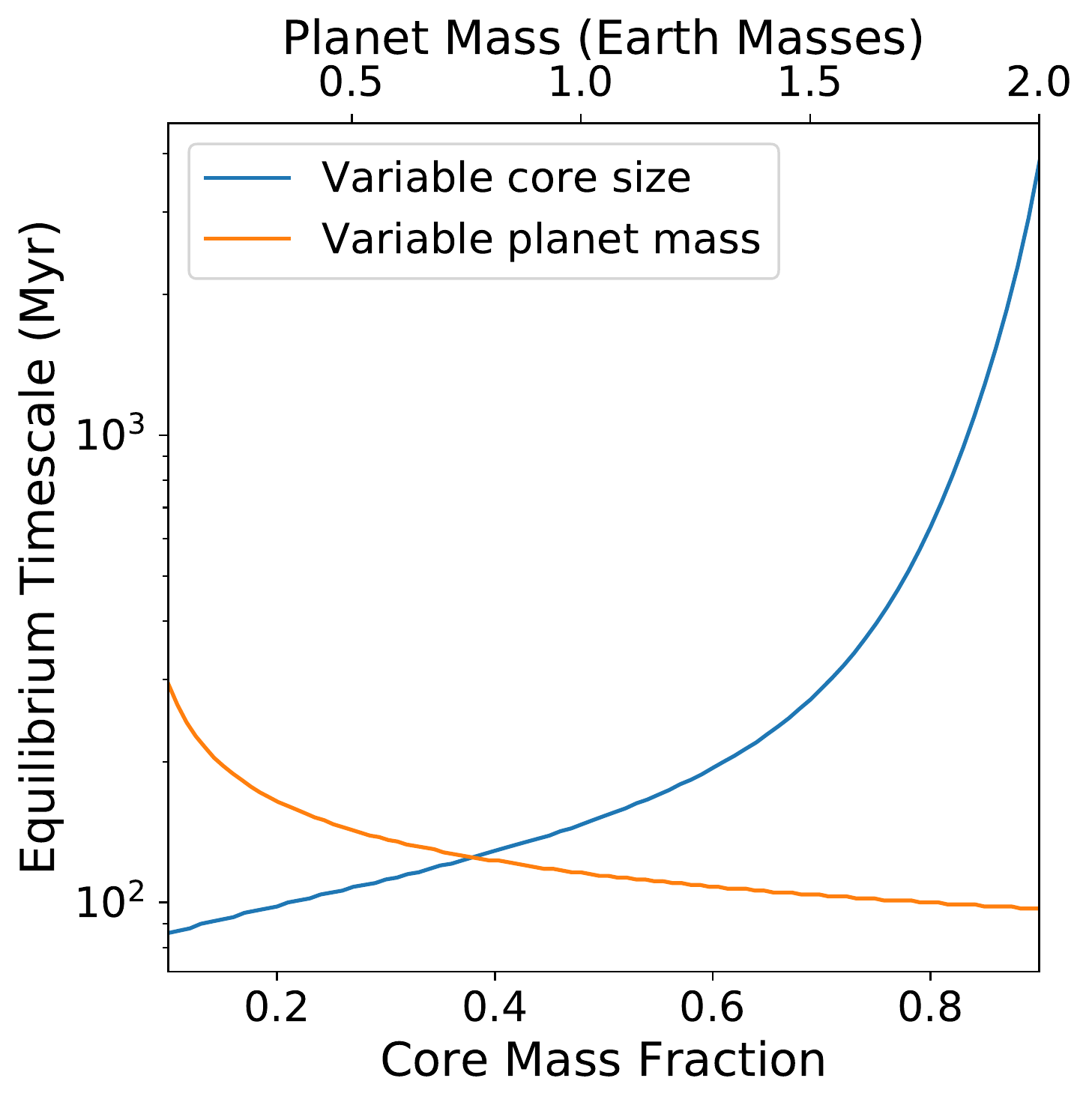}
    \caption{Equilibrium timescale $\tau_\text{CC}$ as a function of core size $f_\text{c}$ and planet mass $M_\text{p}$.}
    \label{fig:35fcMpTimescales}
\end{figure}

\begin{figure}
    \centering
    \includegraphics[width=.5\textwidth]{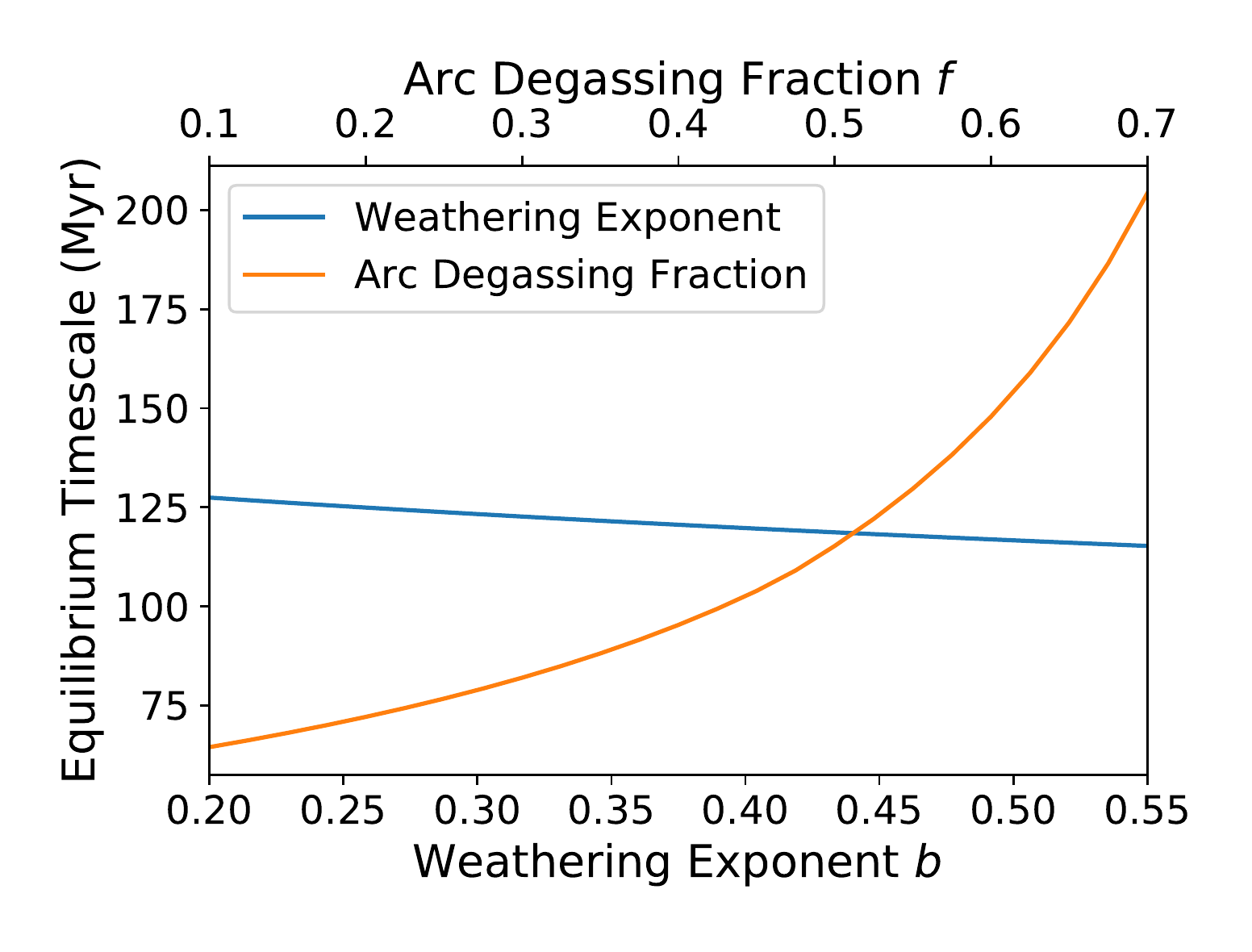}
    \caption{Equilibrium timescale $\tau_\text{CC}$ as a function of weathering exponent $b$ and arc degassing fraction $f$.}
    \label{fig:35CompTauCC}
\end{figure}

Fig. \ref{fig:31PrimaryResultsComparison} shows that after $\sim10^8$ yr, the vast majority of the total carbon budget is stored in the mantle reservoir for any initial distribution of carbon over the various reservoirs, even if the atmosphere is initially carbon-rich. Such an initially CO$_2$-rich atmosphere is thought to be a common consequence of the initial solidification of the crust \citep{Elkins-Tanton2008b}. This subsequently gives rise to surface temperatures well above the moist greenhouse threshold of 340 K, allowing for large amounts of water loss if those temperatures are maintained long enough ($\gtrsim 10^8$ yr) \citep{Kasting1988, Abbott2012}. If the timescale on which the carbon cycle establishes equilibrium between the various carbon reservoirs is $>10^8$ yr, this may be disadvantageous for the development of a long-term, temperate climate with moderate atmospheric CO$_2$ pressures. In this section we explore this \textit{equilibrium timescale} as a function of the interior parameters considered in the previous sections. We also consider this timescale as an indicator of the absolute minimum age an exoplanet must have in order to have a climate which is significantly impacted by a long-term carbon cycle facilitated by plate tectonics.\\
\indent
An estimate for the carbon cycle equilibrium timescale $\tau_\text{CC}$ can be derived by considering the evolution of the carbon reservoirs $R_\text{m}$, $R_\text{k}$ and $R_\text{a}$ for different initial carbon distributions. The initial conditions considered are the same as in Sect. \ref{sec:3.1}. Subsequently, $\tau_\text{CC}$ is chosen to be the time where all three different evolutionary tracks of the three reservoirs $R_i$ all deviate no more than 1\% from the mean reservoir content $\widetilde{R}_\text{i}(t)$ inferred from the three different values of $R_\text{i}(t)$ for the different initial conditions:
\begin{align}
\label{eq:tauCC}
    \tau_\text{CC}:\qquad \left|1-\frac{R_\text{i}(t=\tau_\text{CC})}{\widetilde{R}_{i}(t=\tau_\text{CC})}\right|\leq 0.01.
\end{align}
This expression is evaluated for all three reservoirs, ${R_i\in(R_m, R_k, R_a)}$, and the longest time found is chosen as the equilibrium timescale of the given model. With this definition, one can find $\tau_\text{CC}\approx116$ Myr for the model representing Earth, discussed in Sect. \ref{sec:3.1}.\\
\indent
In the context of mantle abundances of uranium and thorium, we find that $\tau_\text{CC}$ varies moderately (Fig. \ref{fig:35IsotopeTimescales}). Values of $\tau_\text{CC}$ here range from $\sim$ 100 Myr for $f_\text{U}=f_\text{Th}=0.6$ to $\sim$ 120 Myr for $f_\text{U}=f_\text{Th}=2.5$. The lower value of $\tau_\text{CC}$ toward mantles rich in radioactive isotopes is a consequence of the higher plate speeds associated with a hotter mantle, which allows for more rapid sequestration of carbon in the mantle reservoir. The variation in $\tau_\text{CC}$ is moderate due to the fact that equilibrium is already established when the mantle temperature is still close to its initial value, which was chosen to be $T_\text{m0}=2000$ K for all models. Therefore the differences in the thermal budgets induced by radiogenic heating are still small, which propagates into the plate speed and $\tau_\text{CC}$. \\
\indent 
For the core mass fraction $f_c$, the equilibrium timescale increases by over an order of magnitude over the range of core mass fractions considered ($0.1\leq f_\text{c}\leq 0.9$) if the planet mass is fixed at $M_\text{p}=1$ M$_\oplus$ (Fig. \ref{fig:35fcMpTimescales}). This increase is the consequence of two effects, the first being the comparatively low thickness of the mantle $D$ at large core fractions. As the plate speed scales linearly with mantle thickness, planets with a thinner mantle have lower initial plate speeds, as the mantle temperature is initially nearly identical for all models. The second effect is the high surface-volume ratio of the mantle. This allows for efficient mantle cooling at larger core fractions (see also Fig. \ref{fig:33CoreSizeEvolution}), reducing the Rayleigh number and plate speed even further over time. It becomes clear that on planets with large $f_\text{c}$, carbon exchange with the mantle is limited due to the low rates of plate subduction and seafloor spreading.\\
\indent 
As a function of planet mass at constant $f_\text{c}=f_{\text{c}\oplus}=0.3259$, $\tau_\text{CC}$ decreases from $\tau_\text{CC}\approx$ 290 Myr for a $0.1$ M$_\oplus$ planet to $\tau_\text{CC}\approx$ 100 Myr for a $2$ M$_\oplus$ planet, a consequence of the same two effects. More massive planets have mantles which are thicker and hotter, resulting in more vigorous convection, and hence higher plate speeds. This leads to a higher rate of carbon sequestration into the mantle, and hence results in a shorter equilibrium timescale for more massive planets. It becomes clear that carbon exchange with the mantle facilitated by plate tectonics is efficient for a wide range of planetary interiors. \\
\indent In addition to the physical properties of the planet considered, we note that the equilibrium timescale also depends on the parameter values used in our model. An important parameter in this context is the efficiency of carbon sequestration, which is contained in $f$, the fraction of carbon which degasses via arc volcanism upon subduction (see also Appendix \ref{sec:AB}). A high value of $f$ means a large fraction of the subducted carbon is degassed during subduction. This limits the rate at which carbon is sequestered in the mantle, increasing $\tau_\text{CC}$. In our study, we use the value $f=0.5$ (Table \ref{tab:CCTable}), although estimates for $f$ vary from below $0.1$ up to values as large as $0.7$ \citep{Sleep2001, Dasgupta2010, Ague2014, Kelemen2015, Foley2015}. For our model, we find $\tau_\text{CC}$ to vary from $\sim60$ Myr to $\sim200$ Myr within this range, as is shown in Fig. \ref{fig:35CompTauCC}. This variation is significant when compared with the variation in $\tau_{CC}$ found for different radioactive isotope abundances, planet mass and core mass fraction. The uncertainty in the arc volcanism degassing fraction $f$ thus induces a considerable uncertainty in estimates of the equilibrium timescale.\\
\indent Another parameter which could affect the values found for $\tau_\text{CC}$ is the exponent $b$, which controls the direct dependence of continental weathering on the atmospheric CO$_2$ pressure (Eq. \ref{eq:Fweather}). The value of $b=0.55$ used in this study (Table \ref{tab:CCTable}) follows from \cite{Foley2015} and \cite{Driscoll2013}. On one hand, this is considerably larger than values used by \cite{Walker1981} and \cite{Zahnle2002} ($b=0.3$), or by \cite{Berner1991} ($b=0.22$). On the other hand, these values do not imply biological enhancement of weathering \citep[e.g.][]{Schwartzmann1989}. Since biological productivity increases with temperature and atmospheric CO$_2$, a potential biosphere amplifies the CO$_2$-dependence of weathering and thereby climate stability \citep{Caldeira1992, Hoening2020}, which would effectively increase the value of $b$. In any case, a lower value of $b$ weakens the feedback provided by continental weathering, and thereby increases the equilibrium timescale. Figure \ref{fig:35CompTauCC} shows that lower values of the weathering exponent $b$ result in a longer equilibrium timescale, increasing from $\tau_\text{CC}\approx 115$ Myr for $b=0.55$ to $\tau_\text{CC}\approx 127$ Myr for $b=0.2$. This is smaller than the variation found in Fig. \ref{fig:35IsotopeTimescales} for different mantle radioactive isotope abundances, and well bellow the variation in $\tau_\text{CC}$ found as a function of $M_\text{p}$ and $f_\text{c}$. Altogether the effect of $b$ on $\tau_\text{CC}$ remains limited.

    \section{Discussion}
    \label{sec:4}

\subsection{Extrapolation to massive planets}
\label{sec:4.1}
\begin{figure*}
    \centering
    \includegraphics[width=\textwidth]{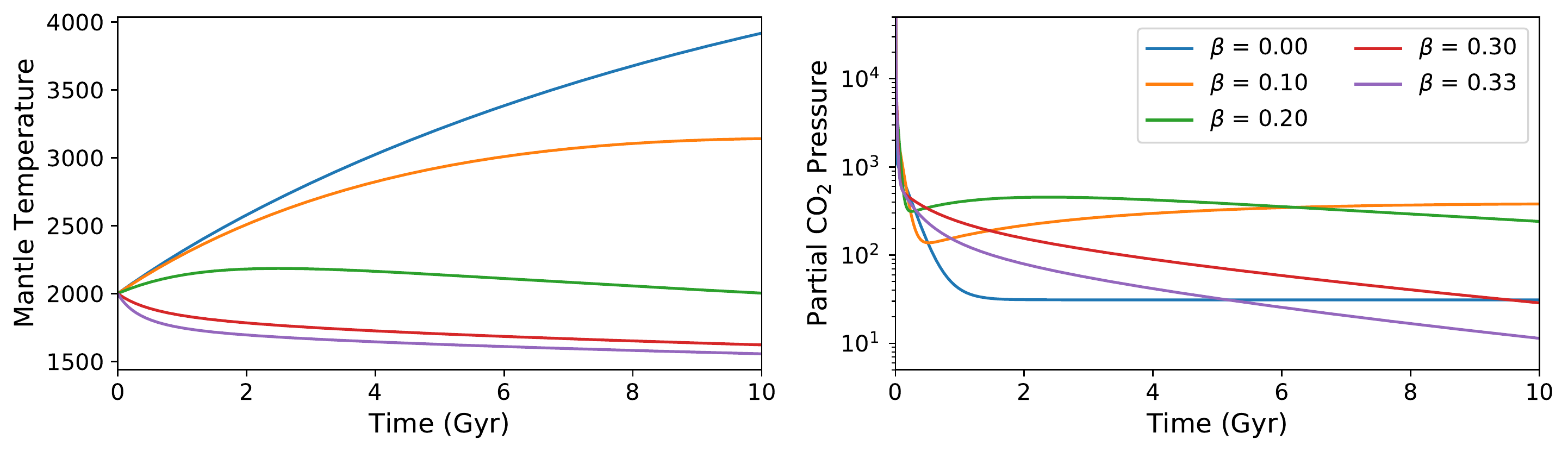}
    \caption{Long-term evolution of the upper mantle temperature $T_\text{m}$ and partial atmospheric CO$_2$ pressure for different values of the Nu-Ra coupling exponent $\beta$.}
    \label{fig:41BetaEvolution}
\end{figure*}
Exploring the effect of planet mass on the carbon cycle, we have assumed the mass-radius relation $R\propto M_\text{p}^{1/3}$, and neglected any effects of the pressure or temperature on the density. However, these assumptions will limit the range of planetary masses that can be meaningfully explored with this model. \cite{Valencia2006} find a mass-radius relation $R\propto M_\text{p}^{\sim0.3}$ for planets with a Mercury-sized core and  $R\propto M_\text{p}^{\sim0.27}$ for an Earth-sized core, with the precise value of the exponent also depending on the equation of state of mantle and core. They also note that for planets less massive than Earth, the effects of pressure and temperature on the density of the mantle and core become less important, resulting in a convergence of the exponent in the mass-radius relationship toward $\frac{1}{3}$, the value associated with constant density. This suggests that for planets less massive than Earth, the constant-density assumption is acceptable.\\
\indent 
Toward more massive planets, we impose that the relative difference between the radius estimates must be smaller than typical uncertainties associated with observational radius measurements. In this context, \cite{Otegi2020} use a relative uncertainty of ${\sigma_R/R\leq8\%}$ as a selection criterion for robust radius measurements of exoplanets. The deviation of ${R\propto M_\text{p}^{1/3}}$ with respect to the ${R\propto M_\text{p}^{0.27}}$ relation exceeds this difference at ${M_\text{p}\approx 3.6}$ M$_\oplus$.\\
\indent 
Besides the effects of pressure on the mean density, the increasing pressure in the interior of massive planets has also effects on the mantle viscosity. Through our work, we assumed the mantle to be isoviscous. However, from literature, conclusions for the viscosity profile throughout the mantle seem to differ for higher-mass planets. Some studies suggest that the isoviscous approximation remains valid for more massive planets \citep{Karato2011}, while others find a strongly increasing viscosity along the same temperature profile throughout the mantle \citep{Stamenkovic2011}. This discrepancy may result from the lack of knowledge of the behavior of mantle rocks under high pressure \citep{Stamenkovic2012}. In general, these studies model the pressure-dependence of the viscosity via extensions of the Arrhenius-type law for purely temperature-dependent viscosity used in our model
\begin{align}
    \nu(T,P)=\nu_0\exp\left(\frac{E^\star + PV_\text{eff}(P)}{R_\text{g}T}\right).
\end{align}
Here $V_\text{eff}(P)$ denotes the effective activation volume, which determines the coupling strength of pressure to viscosity, and is determined by the material properties of the mantle rocks. Altogether this may lead to an increasing viscosity as a function of mantle depth, which can have the formation of a stagnant lid at the core-mantle boundary as a result \citep{Stamenkovic2012}. The formation of such a lower stagnant lid is thought to ensue once the viscosity contrast between the top and bottom of the mantle exceeds values of $10^4$ \citep{Solomatov1995, Solomatov1997}. The viscosity contrast in Earth's mantle is estimated between 2 and 10, well below this range \citep{Yamazaki2001, Paulson2005, Driscoll2014}. On the other hand, planets with masses of $M_\text{p}=5$ M$_\oplus$ and $M_\text{p}=10$ M$_\oplus$ have inferred viscosity contrasts in the range of $\sim 10^5-10^7$ and $\sim 10^{10}-10^{14}$, respectively \citep{Stamenkovic2011}. A range of contrasts is stated as \cite{Stamenkovic2011} also consider variations in the temperature at the core-mantle boundary.\\
\indent 
Altogether there appears to be no straightforward way in estimating the viscosity contrast in a mantle without imposing more detailed assumptions on the mineralogy of the mantles considered. However, the results from \cite{Stamenkovic2011} indicate that an upper boundary of $M_\text{p}=5$ M$_\oplus$ is without doubt too high for our isoviscous mantle model to produce reliable results. To be conservative, we limited ourselves to an upper mass limit of $M_\text{p}=2$ M$_\oplus$ for our parameter space. As temperature and pressure effects are thought to become less important for planets less massive than Earth, we limit the minimum mass considered in our parameter space to $M_\text{p}=0.1$ M$_\oplus$, about the mass of Mars \citep{dePater2015}.

\subsection{The role of coupling exponent $\beta$}
\label{sec:4.2}
The Nu-Ra coupling exponent $\beta$ has been kept fixed at $\beta=\frac{1}{3}$ in this study. This value is based on the assumption that the mantle can be treated as an isoviscous fluid undergoing Rayleigh-Bénard convection, which is also a key assumption underlying our thermal evolution model. $\beta$ is an extremely important parameter as it describes the coupling strength between heat flow and convective vigor and therefore the feedback strength between mantle convection and mantle heat loss. However, the precise value of $\beta$ in the mantles of Earth-like exoplanets is unknown. Therefore, we discuss which values for $\beta$ are possible in the mantles of Earth-like exoplanets. In addition, we show the implications for the long-term evolution of the atmospheric CO$_2$ pressure.\\ 
\indent 
While $\beta=\frac{1}{3}$ follows from theoretical considerations, laboratory experiments suggest $\beta\approx0.3$ to be appropriate for Earth's mantle \citep{Turcotte2014}. However, $\beta$ has been found to vary between 0 and $\frac{1}{3}$ depending on mantle rheology, thermal state and tectonic regime (\citeauthor{Unterborn2015} \citeyear{Unterborn2015}, \citeauthor{Oneil2020} \citeyear{Oneil2020} and references therein). Therefore, $\beta$ could indirectly even be a function of time. For simplicity, however, we restrict our discussion to constant $\beta$, and consider the effects of values within the range of 0 to $\frac{1}{3}$ on mantle temperature $T_\text{m}$ and \PCO\, in Fig. \ref{fig:41BetaEvolution}. Lowering the value of $\beta$ weakens the effect of mantle convection on mantle cooling and plate speed. This means that the mantle temperature increases significantly over time for small $\beta$, up to above 3500 K for $\beta=0$ at $t=10$ Gyr. We note that at these temperatures, widespread mantle melt becomes an important cooling mechanism for the mantle. However, the cooling effects of melting have not been included in our model, but will in reality result in lower mantle temperatures than shown in Fig.~\ref{fig:41BetaEvolution} \citep{Richter1985, Driscoll2014}.\\
\indent
The evolution of the CO$_2$ pressure is also sensitive to the value of $\beta$ due to its dependence on the mean plate speed $v_\text{p}$. The increasing mantle temperature increases the Rayleigh number via the viscosity, resulting in a higher plate speed and hence atmospheric CO$_2$ pressure for the intermediate values $\beta=0.1, 0.2, 0.3$. However, a lower value of $\beta$ also weakens the sensitivity of the plate speed on the Rayleigh number and hence mantle temperature. This explains why we find the highest atmospheric CO$_2$ pressure in Fig. \ref{fig:41BetaEvolution} at intermediate values of $\beta$. $\beta=0$ eliminates the dependence of the plate speed on the Rayleigh number entirely, resulting in the plate speed remaining fixed at its present-day Earth value throughout the entire evolution. It becomes clear that the applicability of our model coupling is limited for smaller values of $\beta$. However, given the similarities to Earth imposed on the planets considered in this study and mantle boundary layer model considered, $\beta=\frac{1}{3}$ appears to be a reasonable choice for this parameter.

\subsection{Plate speed scalings and additional couplings}
\label{sec:4.3}
In this study, we assumed a power-law relation between the mean plate speed $v_\text{p}$ and global Rayleigh number Ra to connect the plate speed to mantle cooling. Furthermore, the plate speed was assumed to scale linearly with the flow speed of the mantle fluid. However, we note that other parametrizations for the plate speed exist, as the evolution of the plate speed does not necessarily have to be dominated by the thermal evolution of the mantle. In addition, the plate speed is not the only variable which couples the carbon cycle to the thermal evolution of the interior. In this section we discuss other parametrizations of the plate speed and their potential effects on our results. In addition, we highlight a few couplings which are not considered explicitly in this study.\\
\indent
In Sect. \ref{sec:3.1} we compared our results to the study of \cite{Foley2015}, where the plate speed was imposed to depend on surface temperature $T_\text{s}$ via the linear relation
\begin{align}
    v_\text{p}(T_\text{s})=a_1-a_2T_\text{s}.
\end{align}
Here $a_1$ and $a_2$ are fitting parameters from earlier work \citep{Foley2014b}. However, we find that the plate speed does not vary significantly over the history of the planet, as depicted in Fig. \ref{fig:31PrimaryResultsComparison}. This different behavior of plate speed evolution with the evolution found with our parametrization is a consequence of the fact that the long-term behavior of the mean plate speed remains poorly understood. One class of models suggests faster plate tectonics in the past as a result of a hotter, more strongly convecting mantle \citep{Hoening2016}. This could explain the observed D/H fractionation on Earth \citep{Kurokawa2018}. Other work suggests that the plate speed remained rather constant as a function of mantle temperature due to the effects of volatile depletion on mantle rheology \citep{Korenaga2003, Korenaga2017}. However, the importance of these effects on the long-term evolution of plate tectonics on Earth is still a topic of debate, making an adequate description for these processes on Earth-like exoplanets challenging.\\
\indent
The long-term behavior of the plate speed in this work is assumed to be controlled by the Rayleigh number via Eq. \ref{eq:plateSpeedScalingLaw}. This expression relies crucially on Eq. \ref{eq:propTo}, the assumption that the plate speed depends linearly on the mantle fluid speed. However, in case a planet has a high lithosphere strengths or a low mantle viscosity, other solutions for the plate speed exist in the so-called sluggish-lid regime. In this case, the plate dynamics is not fully coupled to the dynamics of the mantle fluid, such that $v_\text{p}<u_0$, and the material properties of the lithosphere control the movement of tectonic plates \citep{Crowley2012}. The linear relation can be recovered in the presence of weak lithospheres and low material strength contrasts with respect to the mantle underneath. Altogether a linear relation between plate speed and fluid speed may not hold for exoplanets with mantle compositions and properties that differ significantly from Earth. In addition, a significantly hotter mantle than Earth's may also violate Eq. \ref{eq:propTo}, since the corresponding lower viscosity may decouple the plate motion from the mantle fluid flow \citep{Valencia2007, Berovici2015}. Contrary, it has also been argued that a hotter mantle may ultimately have a higher viscosity when the effects of dehydration on mantle rheology are taken into consideration \citep{Korenaga2003}. Altogether, the precise behavior of plate movement as a function of temperature would require more detailed modeling of the mantle and lithosphere composition and their effects on the mantle viscosity and lithospheric strength with respect to Earth. However, it is important to emphasize that our results would not be affected significantly: The rate of mantle degassing directly depends on the rate at which mantle material enters the source region of partial melt, which is represented by the convection rate $u_0$ rather than by the plate speed $v_\text{p}$ (c.f. Eq. \ref{eq:propTo}). The same applies to the rate of seafloor weathering, which is a function of the crustal production rate, not necessarily of the plate speed \citep[see e.g.][]{Hoening2019b}. Therefore, our results would still hold, even in a case where the plate speed speed does not linearly depend on the convection rate.\\
\indent
Other couplings between the thermal state of the mantle and the climate exist. For example, the decarbonation fraction in subduction zones has been studied as a function of the mantle temperature \citep{Hoening2019b}. Similarly, these authors find declining atmospheric CO$_2$ pressures over time as a consequence of mantle cooling. Furthermore, the mid-ocean ridge melting depth $d_\text{melt}$ has been argued to become larger in response to a hotter mantle \citep{Sleep2001}. This would enhance the amount of ridge degassing at earlier times (Eq. \ref{eq:Fridge}), while gradually decreasing over time as the mantle cools. Altogether the inclusion of the dependence of decarbonization fraction or melting depth on mantle temperature would amplify the declining trend in CO$_2$ pressure over time found in this study.\\
\indent 
The removal of atmospheric CO$_2$ predominantly depends on the weathering rate $F_\text{weather}$. This rate is also indirectly connected to the plate speed. The efficiency of weathering is dependent on the amount of fresh, weatherable rock exposed to the atmosphere via the erosion rate, which is in the long term determined by the amount of uplift caused by plate tectonics \citep[e.g.][]{Kasting2003, Foley2015, Hoening2019a}. Therefore, one can expect weathering fluxes to decline for lower plate speeds. The coupling between continental weathering rate and thermal evolution, however, is left for future studies.

    \section{Implications for exoplanets}
    \label{sec:5}

In this work we aimed to assess the effects of the mantle abundance of radioactive isotopes, core mass fraction and planet mass on the long-term carbon cycles of Earth-like exoplanets. In this section we therefore address the implications of our results for Earth-like exoplanets.\\
\indent
The coupling between mantle cooling and the plate speed suggests gradually decreasing plate speeds over time. As the planet ages and the mantle cools down, we find declining atmospheric CO$_2$ pressure. We note that this decline in CO$_2$ pressure likely amplifies the decline in CO$_2$ pressure in response to the increasing surface temperatures induced by the increasing luminosity of the host star (see Appendix \ref{sec:AC}). This means that if a planet has a carbon cycle facilitated by plate tectonics, one would expect the atmospheres of rocky planets around older G-type stars to be depleted in CO$_2$ with respect to their younger counterparts. An example of such a star which has aged considerably is HD 16417. The age of this G1V-type star has been estimated at $7.0\pm0.4$ Gyr \citep{Baumann2010}, and is known to have at least one Neptune-sized ($M_\text{p}=22.1\pm2.0$~M$_\oplus$) planet \citep{OToole2009}. If such a system were to have a terrestrial planet undergoing carbon cycling as known on Earth, its atmosphere could show an atmospheric CO$_2$ depletion, which could be an indicator of the presence of an Earth-like carbon cycle on such a planet. This could present an independent indicator for the existence of carbon cycling which could be distinguished with for example statistical comparative planetology \citep{Bean2017, Graham2020}.\\
\indent 
Initially high plate speeds resulting from the hot interior allow for rapid exchange of carbon between the planetary surface and mantle, which allows for a rapid transition from a disequilibrium atmosphere to an atmosphere with its CO$_2$ pressure controlled by long-term carbon cycling. This transition occurs within 120 Myr for a wide range of radioactive isotope abundances, core sizes and planet masses. It indicates that if plate tectonics starts on a young, Earth-like exoplanet, its atmospheric CO$_2$ pressure could in principle become fully controlled by the long-term carbon cycle within 120 Myr. Therefore an exoplanet does not have to be old to have its atmospheric CO$_2$ pressure regulated by the long-term carbon cycle. This implies that around younger stars such as HD 141937 ($1.3\pm 0.9$ Gyr old, \citeauthor{Baumann2010} \citeyear{Baumann2010}), one may find planets whose atmospheric CO$_2$ pressure is already controlled by carbon cycling. However, this is only possible if plate tectonics starts early and weathering is efficient at removing atmospheric CO$_2$. \\
\indent
\cite{Unterborn2015} found a spread in the abundance of thorium from 0.6 to 2.5 solar in the atmospheres of a sample of fourteen solar twins and analogs. These stellar atmospheric abundances were derived by means of spectral line fitting. As both thorium and uranium are refractory elements and have a correlated nucleosynthetic origin, we assumed that the abundances of these species in the mantles of terrestrial planets orbiting these stars scale linearly with the abundance of thorium in the atmosphere of the parent star. We found that after 4.5 Gyr, a mantle depleted in radioactive isotopes ($f_\text{Th}=f_\text{U}=0.6$ Earth's mantle abundance) can have a partial CO$_2$ pressure almost an order of magnitude below the partial pressure found on a planet with a mantle rich in radioactive isotopes ($f_\text{Th}=f_\text{U}=2.5$). The abundance of thorium was here found to have a larger effect than the abundance of uranium, in particular at later times. This implies that planets around thorium-rich stars such as HD 102117, HD 141937 and HD 160691 \citep{Unterborn2015} may favor the development of a warmer climate. The higher plate speeds associated with the hotter mantle also facilitate more efficient transport of carbon into and from the mantle. A mantle initially rich in radioactive isotopes also allows the mantle to remain hotter over longer time spans (Fig. \ref{fig:32IsotopeComparison}), and thus maintain these higher plate speeds. This means that efficient carbon cycling is possible over a larger timespan. However, Fig. \ref{fig:35IsotopeTimescales} shows that at earlier times, the equilibrium timescale does not depend as strongly on mantle radioactive isotope abundance, but rather on the initial mantle temperature. Altogether, rocky exoplanets in the plate tectonics regime around thorium- or uranium-rich stars may generally be more efficient in carbon cycling than Earth, in particular at later times.\\
\indent
The core size and mass of a planet also have important consequences for the carbon cycle facilitated by plate tectonics. The rapidly increasing equilibrium timescale in Fig. \ref{fig:35fcMpTimescales} as a function of core size suggests that this mode of carbon recycling is becoming less efficient on planets with a large core. The mantle reservoir starts to decouple from the surface due to the rapid mantle cooling (Fig. \ref{fig:33CoreSizeEvolution}), with the equilibrium timescale approaching 1 Gyr for $f_c\gtrsim 0.8$. Altogether, plate tectonics operates less efficiently on planets with large core sizes; a conclusion consistent with \cite{Noack2014b}, who find that a planet is more likely to be in the stagnant lid regime when the iron core is large, making an Earth-like carbon cycle impossible. For planets that are less massive, we found similar behavior; small planets cool quickly and hence have more sluggish convection in their mantles, while a hotter interior for planets more massive than Earth would allow for faster carbon cycling on these planets.

    \section{Conclusions and outlook}
    \label{sec:6}

In this work we studied the role of the planetary interior in carbon cycling facilitated by plate tectonics and the resulting evolution of the atmospheric CO$_2$ pressure. Interior properties considered are mantle radioactive isotope abundance, planet mass and core mass fraction, parameters which vary significantly for Earth-like exoplanets. In order to assess the effects of these interior parameters on the long-term carbon cycle, we developed a parametrized planetary evolution model which links the thermal evolution of the interior to carbon cycling. The mean plate speed is used as the key coupling variable between the two models. We extrapolated Earth's long-term carbon cycle for a range of these interior parameters. Our findings can be summarized as follows:
\begin{itemize}
    
    \item Including the effects of mantle cooling on the plate speed results in gradually declining atmospheric CO$_2$ pressures over time, due to the gradually decreasing degassing. This decline amounts up to an order of magnitude in \PCO\,over 10 Gyr; a conclusion qualitatively similar to \cite{Hoening2019b}. However, the erosion rate $E_\text{max}$ also declines for lower plate speeds, and thus may cause the continental weathering rate to decline in response to lower plate speeds. Feedback between the plate speed and these variables could be important to consider in future studies, and assess whether this feedback stabilizes the atmospheric partial CO$_2$ pressure against decreased degassing.
    
    \item A long-term carbon cycle driven by plate tectonics could operate efficiently on planets with amounts of radiogenic heating in their mantles different from Earth. However, planets with their mantles enriched in radioactive isotopes with respect to Earth, may favor the development of warmer climates resulting from a more CO$_2$ rich atmosphere. This is in particular the case for a planet with a higher thorium abundance. In addition, the carbon cycle operates more efficiently on planets rich in radioactive isotopes, motivating the characterization of planetary systems around stars whose atmospheres are rich in thorium or uranium.
    
    \item A long-term carbon cycle regulated by plate tectonics may not be feasible for planets with core mass fractions $f_c\gtrsim 0.8$. This can mainly be attributed to the decrease of the Rayleigh number over time in response to fast mantle cooling. This effect reduces the effectiveness of plate tectonics in sequestering carbon in the mantle through subduction. Altogether this result emphasizes the importance of high-precision mass-radius measurements of Earth-sized exoplanets, which can help in constraining the core sizes of these exoplanets.
    
    \item Planets with plate tectonics may favor higher atmospheric CO$_2$ pressure for higher total mass or smaller core fractions than Earth's. Both result in a hotter mantle with a higher Rayleigh number, which promotes degassing through higher plate speeds. Future studies should focus on extrapolating these results toward planet masses above 2 M$_\oplus$, which requires the incorporation of the pressure-dependence of the mantle viscosity and density in the thermal evolution model.
    
    \item Carbon cycling facilitated by plate tectonics can regulate the atmospheric CO$_2$ pressure from disequilibrium values comparatively fast; an equilibrium timescale between 100 and 200 Myr was found throughout most of the parameter space explored. This means that provided plate tectonic initiates, an exoplanet does not necessarily have to be old to have an atmosphere significantly altered by a long-term carbon cycle. 
\end{itemize}
Altogether we find that the thermal evolution of the planetary interior has a significant effect on the evolution of the atmospheric CO$_2$ pressure. However, for a large fraction of the parameter space explored in this study, this does not significantly impede the efficiency at which the long-term carbon cycle is able to regulate the atmospheric CO$_2$ pressure.

    \section*{Acknowledgements}
    We would like to thank Bradford Foley for his help in benchmarking the carbon cycle model, and an anonymous reviewer for constructive comments. D.H. has been supported through the NWA StartImpuls.

%
%

\bibliographystyle{aa.bst}
\bibliography{refdata.bib}

\begin{appendix}
\section{Thermal evolution model}
\label{sec:AA}
\begin{figure}
    \centering
    \includegraphics[width=.49\textwidth]{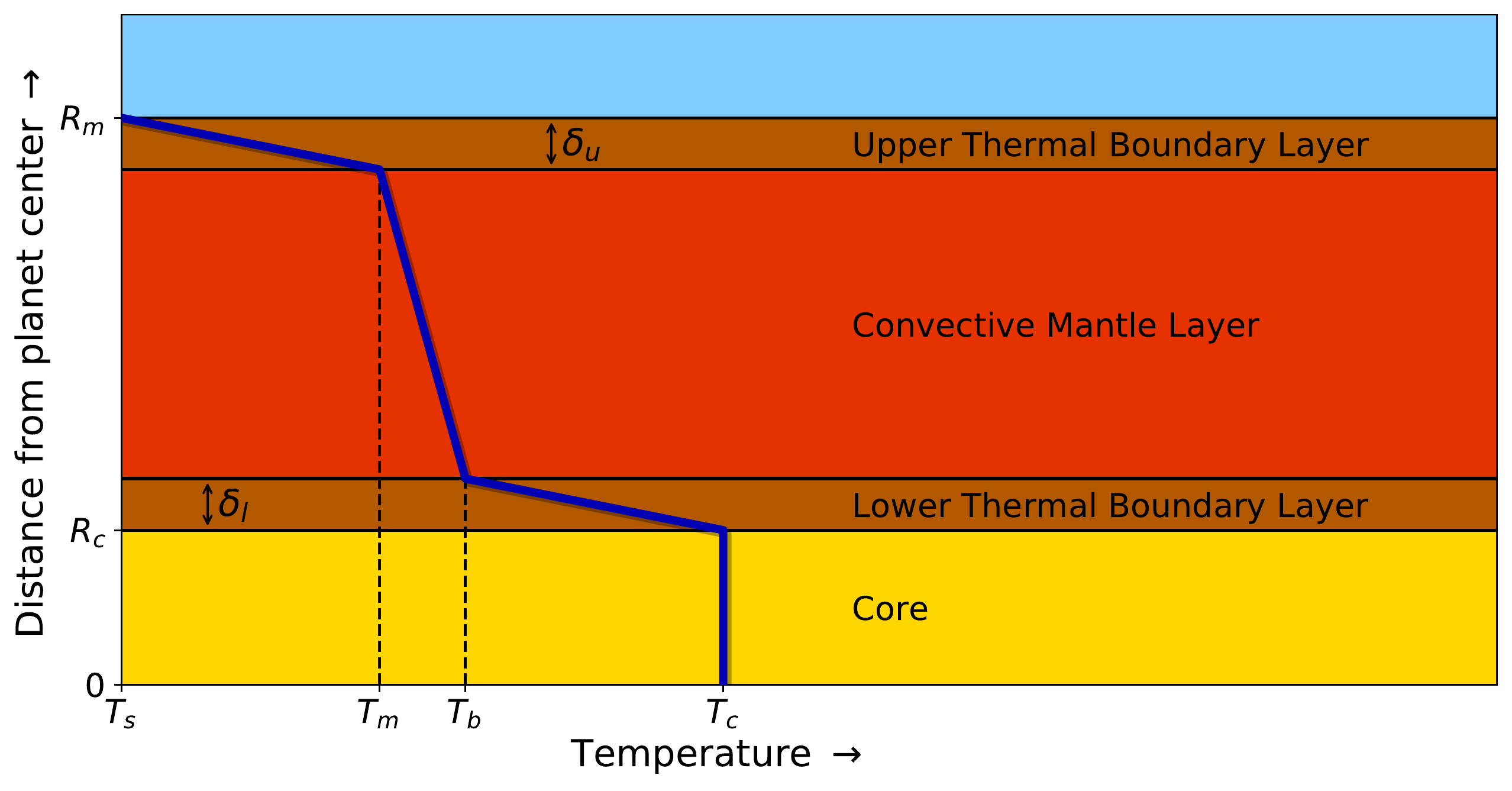}
    \caption{Overview of the temperature profile (blue) and thermal structure considered in the thermal evolution model}
    \label{fig:AATemperatureProfileOverview}
\end{figure}

The evolution for the mantle temperature $T_\text{m}$ and core temperature $T_\text{c}$ are calculated by integrating the following energy conservation equations for the mantle and core over time \citep[e.g.][]{Schubert2001}:
\begin{align}
    \label{eq:eConMan}
      V_\text{m}\,\rho_\text{m}\,c_\text{m} \parder{T_\text{m}}{t}&= V_\text{m}\,H(t)-A_\text{m}\,q_\text{u}+A_\text{c}\,q_\text{l}\\
      V_\text{c}\,\rho_\text{c}\,c_\text{c}\parder{T_\text{c}}{t}&=-A_\text{c}\,q_\text{l}.
\end{align}
Here, $V_\text{m}$ and $A_\text{m}$ denote the volume and outer surface area of the mantle, while $\rho_\text{m}$ and $c_\text{m}$ its density and specific heat capacity. Furthermore, quantities with subscript c denote the same quantities for the core.\\
\indent
$H(t)$ denotes the volumetric specific radiogenic heating per unit mantle volume provided by the radioactive isotopes $^{235}$U, $^{238}$U, $^{232}$Th and $^{40}$K,
\begin{align}
    H(t)=Q_0\sum\limits^\text{isotopes}_iq_i\exp\left(-\frac{(t-t_\oplus)\ln (2)}{t_{\text{h},i}}\right).
\end{align}
Here $Q_0$ denotes the total present-day Earth heat production per unit volume. $t_\oplus$ and $t_{\text{h},i}$ denote the age of Earth and half-life of radioactive isotope $i$, respectively. $q_i$ is the percentage of $Q_0$ provided by radioactive isotope $i$.\\
\indent
In addition, $q_\text{u}$ and $q_\text{l}$ denote the heat fluxes through the upper and lower thermal boundary layer, respectively:
\begin{align}
    q_\text{u}=k\frac{T_\text{m}-T_\text{s}}{\delta_\text{u}};\qquad q_\text{l}=k\frac{T_\text{c}-T_\text{b}}{\delta_\text{l}}.
\end{align}
Here $k$ denotes the mantle thermal conductivity, $T_\text{s}$ the surface temperature and $T_\text{b}$ the temperature at the top of the lower thermal boundary layer (see Fig. \ref{fig:AATemperatureProfileOverview}). $T_\text{b}$ is related to $T_\text{m}$ via the linearized adiabatic temperature profile
\begin{align}
\label{eq:TB}
    T_\text{b}=T_\text{m}\left(1+\frac{\alpha g (D-2\delta_\text{u})}{c
    _\text{m}}\right).
\end{align}
Here, $\alpha$ denotes the thermal expansivity, $g$ the mean gravitational acceleration throughout the mantle and $D=R_\text{m}-R_\text{c}$ the mantle thickness. Furthermore, we assume the thickness of the lower thermal boundary layer $\delta_\text{l}\eqsim \delta_\text{u}$. Furthermore, $\delta_\text{u}$ is dictated by a local instability criterion \citep{Stevenson1983} where we impose that the local Rayleigh number Ra$_\text{u}$ of the thermal boundary layer is equal to the critical Rayleigh number:  
\begin{align}
    \text{Ra}_\text{u}=\frac{g\alpha\rho_\text{m}^2c_\text{m}(T_\text{m}-T_\text{s})\delta_\text{u}^3}{k\eta(T_\text{m})}\equiv \text{Ra}_\text{cr}.
\end{align}
Here, $\eta(T_\text{m})$ denotes the mantle viscosity. One thus finds for $\delta_\text{u}$:
\begin{align}
    \delta_\text{u}=\left(\frac{k\eta(T_\text{m})}{g\alpha\rho_\text{m}^2c_\text{m}(T_\text{m}-T_\text{s})}\text{Ra}_\text{cr}\right)^{1/3}.
\end{align}
For general Nu-Ra coupling exponent $\beta$ (see Sect. \ref{sec:2.3}), the above relation can be written as \citep{Driscoll2014}
\begin{align}
\label{eq:generalDeltaU}
\delta_\text{u}=D\left(\frac{k\eta(T_\text{m})}{g\alpha\rho_\text{m}^2c_\text{m}(T_\text{m}-T_\text{s})D^3}\text{Ra}_\text{cr}\right)^{\beta}.
\end{align}
Furthermore, we assume the mantle to be isoviscous, with its viscosity being given by the Arrhenius law
\begin{align}
\label{eq:viscosity}
    \eta(T_\text{m})=\eta_\oplus\exp\left(\frac{E^\star}{R_g}\left[\frac{1}{T_\text{m}}-\frac{1}{T_{\text{m}\oplus}}\right]\right),
\end{align}
where $E^\star$ denotes the activation energy and $R_\text{g}$ is the ideal gas constant. $\eta_\oplus$ denotes a reference viscosity at reference temperature $T_{\text{m},\oplus}$, for which we choose values representative for the present-day Earth upper mantle. An overview of all values used for the parameters introduced in this section is presented in Table~\ref{tab:TETable}.

\section{Carbon cycle model}
\label{sec:AB}
The carbon cycle model we use closely follows the model considered by \cite{Foley2015}, who developed this model to investigate the response of the carbon cycle to different amounts of land coverage and carbon participating in the cycle. We here present a technical overview of this model, while we refer to \cite{Foley2015} for a full motivation and explanation of the model parameters used.\\
\indent
We consider the partition of a constant total carbon budget $R_\text{tot}$ participating in the long-term carbon cycle over various reservoirs for the atmosphere $R_\text{a}$, ocean $R_\text{o}$, oceanic crust $R_\text{k}$ and mantle $R_\text{m}$. $R_i$ here denotes the amount of carbon present in reservoir $i$ in moles.
\begin{align}
    \der{(R_\text{a}+R_\text{o})}{t}&=F_\text{ridge}+F_\text{arc}-\frac{1}{2}F_\text{weather}-F_\text{sfw},\label{eq:Rs}\\
    \der{R_\text{k}}{t}&=\frac{1}{2}F_\text{weather}+F_\text{sfw}-F_\text{sub},\label{eq:Rk}\\
    \der{R_\text{m}}{t}&=(1-f)F_\text{sub}-F_\text{ridge}.
\end{align}
Here, $F_\text{ridge}$ denotes the carbon flux (in moles per unit time) due to degassing of carbon from the mantle at mid-oceanic ridges. $F_\text{arc}=fF_\text{sub}$ is the degassing of CO$_2$ from subduction zones via arc volcanism, and $F_\text{sub}$ denotes the carbon flux into subduction zones, with $f$ being the fraction of the carbon which is degassed during subduction. We note that $f$ is not well constrained \citep{Sleep2001, Dasgupta2010, Ague2014, Foley2015} and depends on the thermal structure of subduction zones, which evolves over time \citep{Johnston2011, Hoening2019b}. However, for simplicity, we keep $f$ fixed at value of 0.5 throughout this work. The factor of $\frac{1}{2}$ in equations \ref{eq:Rs} and \ref{eq:Rk} results from the partial release of carbon back into the atmosphere upon the deposition of carbon on the seafloor in the form of carbonate rocks \citep{Kasting2003}.\\
\indent
The partition of carbon between the atmosphere and ocean is assumed to be instantaneous, and dictated by Henry's law, which can be written as
\begin{align}
\label{eq:Henry}
    P_{\CO}=k_\text{c}x_\text{c}=k_\text{c}\frac{R_\text{o}}{M_{\H_2\O}+R_\text{o}}.
\end{align}
Here, \PCO\, denotes the partial atmospheric CO$_2$ pressure, and $k_c$ the solubility of CO$_2$ in water. \PCO\, can be written in terms of $R_\text{a}$ via
\begin{align}
\label{eq:PCO2}
     P_{\CO}=\frac{R_\text{a}m_{\CO}g}{A_\text{m}}.
\end{align}
$m_{\CO}$ denotes the molar mass of $\CO$, $g$ the surface gravity, and $A_\text{m}$ denotes the surface area of the planet. Combination of equations \ref{eq:Henry} and \ref{eq:PCO2} allows one to infer $R_\text{a}$ and $R_\text{o}$ at each timestep, with $R_\text{a}+R_\text{o}$ being calculated at each timestep via equation \ref{eq:Rs}.\\
\indent
For continental weathering we distinguish between two weathering regimes: a kinetically-limited regime and a supply-limited regime \citep{West2012, Foley2015}. However, we note the existence of more recent treatments for the continental weathering rate which for example distinguish an additional "runoff-limited" regime \citep[][]{Graham2020, Hakim2020}. In the kinetically-limited regime, the weathering rate is limited by the weathering reaction rate and hence depends on surface temperature $T_\text{s}$ and partial atmospheric CO$_2$ pressure \PCO. However, if the weathering rate becomes very high, the weathering rate is limited by the rate at which weatherable rock is exposed to the atmosphere. We use a parametrization which captures the behavior of continental weathering in both regimes, following \citep{West2012, Foley2015},
\begin{multline}
\label{eq:Fweather}
    F_\text{weather}=F_{\text{w}_\text{s}}\left[1-\exp\left(-\frac{F_{\text{w}\oplus}f_\text{land}}{F_{\text{w}_\text{s}}f_{\text{land}\oplus}}\left(\frac{P_\text{sat}}{P_{\text{sat}\oplus}}\right)^a\right.\right.\\
    \left.\left.\cdot\left(\frac{P_\text{CO$_2$}}{P_{\text{CO$_2$},\oplus}}\right)^b\exp\left[\frac{E_\text{a}}{R_\text{g}}\left(\frac{1}{T_{\text{s}\oplus}}-\frac{1}{T_\text{s}}\right)\right]\right)\right].
\end{multline}
Here, $F_{\text{w}\oplus}$ denotes the present-day Earth weathering rate, $f_\text{land}$ the fraction of the surface area covered by land, $E_\text{a}$ the activation energy for the weathering reaction, $R_\text{g}$ the ideal gas constant, and $T_\text{s}$ the average surface temperature. Values denoted with $\oplus$-subscript indicate the respective parameter for present-day Earth. Furthermore, $a$ and $b$ are constants, and $P_\text{sat}$ is the saturation pressure for water vapor, given by
\begin{align}
    P_\text{sat}=P_\text{sat0}\exp\left[-\frac{m_\text{w}L_\text{w}}{R_\text{g}}\left(\frac{1}{T_\text{s}}-\frac{1}{T_\text{sat0}}\right)\right],
\end{align}
where $P_\text{sat0}$ denotes the reference saturation pressure for water vapor at temperature $T_\text{sat0}$.  $m_\text{w}$ denotes the molar mass of water and $L_\text{w}$ the latent heat of water. The continental weathering rate is scaled with a saturation vapor pressure term in order to account for variations in runoff for different $T_s$ \citep{Driscoll2013, Foley2015}.
\\
\indent
When weathering becomes supply-limited, the dependence on $T_\text{s}$ and \PCO disappears. Different parametrizations describing the continental weathering rate in the supply-limited regime $F_{\text{w}_\text{s}}$ exist \citep[e.g.][]{Riebe2004, Foley2015}. In this work, we use the expression derived by \cite{Foley2015},
\begin{align}
\label{eq:Fsup}
    F_{\text{w}_\text{s}}=\frac{A_\text{m}f_\text{land}E_\text{max}\rho_\text{r}\chi_\text{cc}}{\bar{m}_\text{cc}}.
\end{align}
Here $E_\text{max}$ denotes the globally averaged erosion rate, for which we follow \cite{Foley2015} and use an upper bound value which is kept constant over time. Furthermore $\rho_\text{r}$ denotes the average regolith density, $\chi_{cc}$ is the fraction of reactable cations in the surface rock, and $\bar{m}_\text{cc}$ the average molar mass of molecules in the rock which participate in silicate weathering.\\
\indent For seafloor weathering, we use a parametrization which depends on $P_\text{CO$_2$}$ and plate speed $v_\text{p}$, although we note that different parametrizations exist, for example in terms of the ocean floor temperature and ocean pH \citep{Coogan2015, KrissansenTotton2017}. In case of seafloor weathering dependent on $P_\text{CO$_2$}$ and $v_\text{p}$, one can write the scaling law \citep{Sleep2001, Mills2014, Foley2015}
\begin{align}
\label{eq:Fsfw}
    F_\text{sfw}=F_{\text{sfw}\oplus}\left(\frac{v_\text{p}}{v_{\text{p}\oplus}}\right)\left(\frac{P_\text{CO$_2$}}{P_{\text{CO$_2$}\oplus}}\right)^\alpha.
\end{align}
$F_{\text{sfw}\oplus}$, $v_{\text{p}\oplus}$ and $P_{\text{CO$_2$}\oplus}$ here denote the present-day Earth seafloor weathering rate, plate speed and partial CO$_2$ pressure, respectively.\\ \indent
The subduction flux $F_\text{sub}$, can be written in terms of the surface area of seafloor entering subduction zones per unit time multiplied by the density of carbon on the seafloor,
\begin{align}
\label{eq:Fsub}
    F_\text{sub}=v_\text{p}L\frac{R_\text{k}}{(1-f_\text{land})A_\text{m}},
\end{align}
where $L$ denotes the total length of subduction zones.\\
\indent
Lastly, the degassing flux at mid-oceanic ridges, $F_\text{ridge}$ is given by
\begin{align}
\label{eq:Fridge}
    F_\text{ridge}=f_\text{d}\frac{R_\text{m}}{V_\text{m}}2v_\text{p}Ld_\text{melt}.
\end{align}
Here, $f_\text{d}$ denotes the fraction of the carbon in the upwelling mantle material that is released into the atmosphere, and $d_\text{melt}$ the scaling depth at which mantle rocks start to melt due to depressurization \citep{Schubert2001}.\\
\indent
Lastly we relate $P_\text{CO$_2$}$ to $T_\text{s}$. For this purpose we use the parametrization derived by \cite{Walker1981}.
\begin{align}
\label{eq:surfTemp}
    T_\text{s} = T_{\text{s}\oplus}+2(T_\text{e}-T_{\text{e}\oplus})+4.6\left[\left(\frac{P_\text{CO$_2$}}{P_{\text{CO$_2$},\oplus}}\right)^{0.346}-1\right].
\end{align}
Here, $T_\text{e}$ the effective temperature and $T_{\text{e}\oplus}$ the present-day Earth effective temperature. The effective temperature of a planet is given by \citep{dePater2015}
\begin{align}
\label{eq:Te}
    T_\text{e}=\left(\frac{S_\text{irr}(1-A)}{4\sigma}\right)^{1/4},
\end{align}
where $S_\text{irr}$ denotes the solar flux, $A$ the planetary albedo, and $\sigma$ the Stefan-Boltzmann constant. We note that $S_\text{irr}$ is a function of time, depending on the evolution of the host star and migration of the planet's orbit \citep{Gough1981, Lubow2010}. However, initial model runs revealed that the Sun has a very strong influence on the evolution of atmospheric CO$_2$ as its evolution provides a strong external forcing on the surface temperature and therefore weathering. For simplicity, we therefore fixed the solar flux at its present-day Earth value $S_\oplus=1360\,\unit{W}{}\unit{m}{-2}$ throughout this work, while the effects of variable $S_\text{irr}$ are considered in Appendix \ref{sec:AC}. An overview of all model parameters associated with the carbon cycle model is presented in Table~\ref{tab:CCTable}.

\section{Effects of stellar evolution on atmospheric CO$_2$ content}
\label{sec:AC}
\begin{figure}
    \centering
    \includegraphics[width=.49\textwidth]{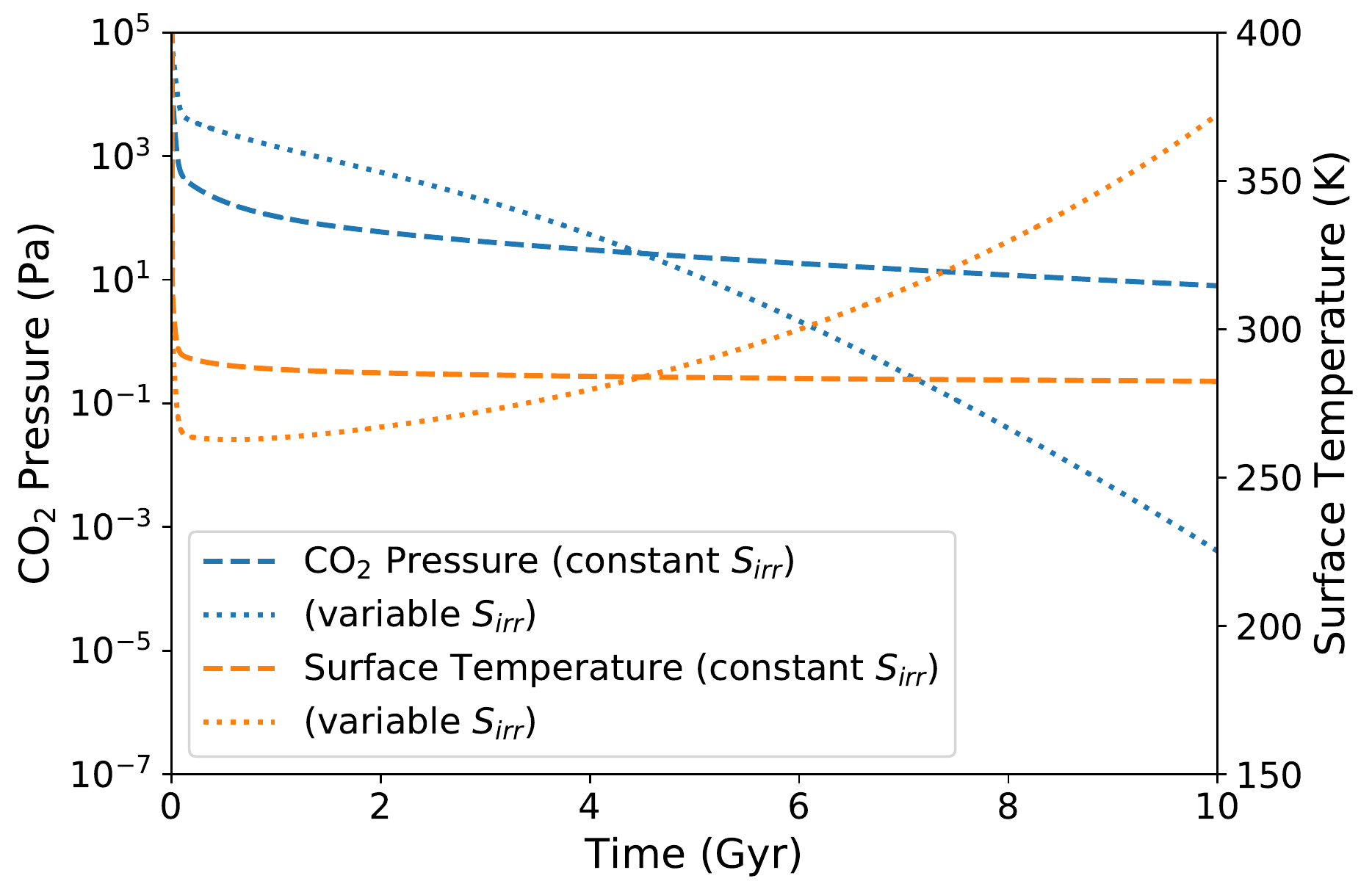}
    \caption{Evolution of the atmospheric partial CO$_2$ pressure and mean surface temperature for a case where $S_\text{irr}=S_\odot$ and a case where $S_\text{irr}$ evolves according to Equation \eqref{eq:varSol}.}
    \label{fig:ACVarSolComparison}
\end{figure}

The cooling of the interior is shown in Sect. \ref{sec:3.1} to have important effects on the long-term behavior of the carbon reservoirs before equilibrium is established. However, another process which becomes relevant on timescales longer than $t\sim10^9$ yr is the gradual increase of the stellar luminosity $S_\text{irr}$ \citep{Gough1981, Foley2015}. Therefore, we consider the effects both the thermal evolution and stellar evolution on the evolution of the atmospheric $\CO$ pressure, and assess their relative importance for driving the long-term evolution of the atmospheric $\CO$ pressure.\\
\indent
In this analysis, we restrict ourselves to Sun-like stars, such that the evolution of the stellar irradiation is given by the following parametrization derived by \cite{Gough1981} for the Sun:
\begin{align}
\label{eq:varSol}
    S_\text{irr}=\frac{S_\odot}{1+\frac{2}{5}\left(1-\frac{t}{t_\odot}\right)}.
\end{align}
We note however that the precise evolution of stellar irradiation over time is a function of both stellar spectral type and planetary migration. Incorporating Equation \eqref{eq:varSol} into \eqref{eq:Te}, we show the evolution of the partial atmospheric CO$_2$ pressure $P_{\C\O_2}$ and surface temperature $T_s$ in Fig. \ref{fig:ACVarSolComparison}. We here compare a case where $S_\text{irr}$ is fixed at its present-day value with a case where $S_\text{irr}$ is allowed to evolve over time. We set $R_\text{s0}=R_\text{tot}$ and $R_\text{m0}=R_\text{k0}=0$ as initial distribution of carbon for all these models. Furthermore, we use the same parameter values and initial conditions as for the results in Fig. \ref{fig:31PrimaryResultsComparison}. Though we only focus on the long-term behavior, it should be noted that the converging behavior toward equilibrium for different initial distributions of carbon remains similar when the evolution of the Sun is considered.\\
\indent
Figure \ref{fig:ACVarSolComparison} reveals two effects of the evolution of the Sun and interior on the planetary evolution. First of all, the evolution of the Sun makes a crucial difference for the evolution of $P_{\C\O_2}$, with atmospheric CO$_2$ pressure dropping from values $\sim10^4$ Pa at $t\approx 1$ Gyr to $P_{\C\O_2}\sim 10^{-4}$ Pa at $t\sim 10$ Gyr. This decrease in atmospheric $\CO$ results from the fact that the brightening of the Sun over time increases the surface temperature. A higher surface temperature in turn increases the amount of continental weathering, given by Equation \ref{eq:Fweather}, which subsequently results in depletion of atmospheric CO$_2$, dampening the effect of the Sun on the surface temperature. However, this dampening effect by the carbon cycle appears to be insufficient to sustain surface temperatures above the freezing point of water ($T_s<273$ K) at earlier times ($t\lesssim 3$ Gyr), and below the moist greenhouse threshold ($T_s>340$ K) at later times ($t\gtrsim8.5$ Gyr). The second message of Fig. \ref{fig:ACVarSolComparison} is with respect to the coupled model; the gradually declining degassing due to the cooler interior may actually help in depleting the atmosphere of CO$_2$ to offset the gradually brightening Sun, while helping in sustaining a more CO$_2$-rich atmosphere at earlier times. However, it appears that this stabilizing effect is only very minor, and other mechanisms are required to mitigate the effects of the faint Sun at earlier times, and the bright Sun at later times, as is also noted by for example \cite{Sleep2001}.

\onecolumn
\section{Parameter tables}
\subsection{Thermal evolution model}
\begin{longtable}{cp{0.6\textwidth}rl}
    \hline
         \textbf{Parameter}& \textbf{Name} &\textbf{Value} & \textbf{Unit} \\\hline\hline
$A_{\text{c}\oplus}$       & Earth's core surface area$^{(2)}$                     &   $1.5\e{14}$ &   $\unit{m}{2}$\\
$A_{\text{m}\oplus}$       & Earth's mantle surface area$^{(2)}$                     &   $5.1\e{14}$ &   $\unit{m}{2}$\\
$\alpha$     & Mantle mean thermal expansivity$^{(2)}$   &   $2\e{-5}$   &   $\unit{K}{-1}$  \\
$\beta$     & Nu-Ra relation exponent            & $1/3$         & -\\           
$c_\text{c}$       & Core mean specific heat$^{(5)}$           & $800$       &   $\unit{J}{}\unit{kg}{-1}\unit{K}{-1}$\\
$c_\text{m}$       & Mantle mean specific heat$^{(5)}$         & $1100$      &   $\unit{J}{}\unit{kg}{-1}\unit{K}{-1}$\\
$D_\oplus$         & Earth's mantle thickness$^{(4)}$      & $2.92\e{6}$   & $\unit{m}{}$ \\
$E^\star$   & Activation energy for subsolidus creep deformation$^{(2)}$ & $3\e{5}$ & $\unit{J}{}\unit{mol}{-1}$\\
$g$         & Gravitational acceleration$^{(2)}$        & $9.81$        & $\unit{m}{}\unit{s}{-2}$\\
$k$         & Mantle mean thermal conductivity$^{(2)}$  &   $4$         &   $\unit{W}{}\unit{m}{-1}\unit{K}{-1}$\\
$\eta_\oplus$    & Present-day Earth viscosity$^{(2)}$       & $10^{21}$    & $\unit{Pa}{}\unit{s}{}$\\ 
$Q_0$       & Present-day radiogenic heat production$^{(1)}$ & $3.31\e{-8}$ & $\unit{W}{}\unit{m}{-3}$\\
$q_{^{238}\text{U}}$ & Present-day heat production contribution from $^{238}$U$^{(1)}$ & $24.285$ & \% \\
$q_{^{235}\text{U}}$ & Present-day heat production contribution from $^{235}$U$^{(1)}$ & $0.1467$ & \% \\
$q_{^{232}\text{Th}}$ & Present-day heat production contribution from $^{232}$Th$^{(1)}$ & $74.752$ & \% \\
$q_{^{40}\text{K}}$ & Present-day heat production contribution from $^{40}$K$^{(1)}$ & $0.8063$ & \% \\
$R_\text{g}$         & Gas constant$^{(2)}$                      &   $8.314$     &   $\unit{J}{}\unit{mol}{-1}\unit{K}{-1}$\\
$\text{Ra}_\text{cr}$   & Critical Rayleigh number$^{(2)}$      & $450$       & - \\
$\text{Ra}_\oplus$      & Present-day Earth Rayleigh number$^{(5)}$ & $4.26\e{7}$& - \\
$\rho_\text{c}$                & Core mean density$^{(5)}$              &   $11449$    &   $\unit{kg}{}\unit{m}{-3}$\\
$\rho_\text{m}$                & Mantle mean density$^{(5)}$             &  $4424$    &   $\unit{kg}{}\unit{m}{-3}$\\
$t_\oplus$   & Age of Earth$^{(2)}$ & $4.5$     & $\unit{Gyr}{}$\\
$t_{h,^{238}\text{U}}$ & $^{238}$U half-life$^{(3)}$ & $4.468$ & $\unit{Gyr}{}$ \\
$t_{h,^{235}\text{U}}$ & $^{235}$U half-life$^{(3)}$ & $0.7038$ & $\unit{Gyr}{}$ \\
$t_{h,^{232}\text{Th}}$ & $^{232}$Th half-life$^{(3)}$ & $14.05$ & $\unit{Gyr}{}$ \\
$t_{h,^{40}\text{K}}$ & $^{40}$K half-life$^{(3)}$ & $1.277$ & $\unit{Gyr}{}$ \\
$T_{\text{m}\oplus}$   & Present-day Earth upper mantle temperature$^{(2)}$ & $1650$     & $\unit{K}{}$\\
$V_{\text{c}\oplus}$           & Earth's core volume$^{(2)}$             & $1.7\e{20}$ &   $\unit{m}{3}$\\
$V_{\text{m}\oplus}$           & Earth's mantle volume$^{(2)}$           & $9.1\e{20}$ &   $\unit{m}{3}$\\ \hline
\captionsetup{width=.9\textwidth}
    \caption{Input parameters used for the mantle model of Sect. \ref{sec:2.1} and Appendix \ref{sec:AA}. Parameter values are taken from \cite{Korenaga2008}$^{(1)}$, \cite{Hoening2019b}$^{(2)}$, \cite{Turcotte2014}$^{(3)}$, \cite{Driscoll2014}$^{(4)}$ or calculated in this work $^{(5)}$}
    \label{tab:TETable}
    \end{longtable}

\newpage
\subsection{Carbon cycle model}
    \begin{longtable}{cp{0.6\textwidth}rl}
    \hline
         \textbf{Parameter}& \textbf{Name} &\textbf{Value} & \textbf{Unit} \\\hline\hline
$a$                 & Saturation vapor pressure scaling law exponent                    &$0.3$&-\\
$a_1$               & Plate speed coefficient 1                                         &$8.0592$&$\unit{cm}{}\unit{yr}{-1}$\\
$a_2$               & Plate speed coefficient 2                                         &$0.0107$&$\unit{cm}{}\unit{yr}{-1}\unit{K}{-1}$\\
$A$                 & Average planetary albedo                                          &$0.31$&-\\
$A_\text{m}$               & Mantle surface area                                     &$5.1\e{14}$&$\unit{m}{2}$\\
$\alpha$            & Partial CO$_2$ pressure $F_\text{sfw}$ scaling law exponent       &$0.25$&-\\
$b$                 & Partial CO$_2$ pressure $F_\text{weather}$ scaling law exponent   &$0.55$&-\\
$d_\text{melt}$     & Mid-ocean ridge melting depth                                     &$70$&$\unit{km}{}$\\
$E_\text{a}$               & Weathering reaction activation energy                             &$42\e{3}$&$\unit{J}{}\unit{mol}{-1}$\\
$E_\text{max}$      & Upper bound on erosion rate                                       &$10^{-3}$&$\unit{m}{}\unit{yr}{-1}$\\
$f$                 & Subduction zone CO$_2$ degassing fraction                         &$0.5$&-\\
$f_\text{d}$               & Mantle CO$_2$ degassing fraction at mid ocean ridges              &$0.32$&-\\
$f_{\text{land}\oplus}$ & Present-day Earth land fraction                                   &$0.3$&-\\
$F_{\text{sfw}\oplus}$  & Present-day Earth seafloor weathering flux                        &$1.75\e{18}$&$\unit{mol}{}\unit{Myr}{-1}$\\
$F_{\text{w}\oplus}$           & Present-day Earth continental weathering flux                     &$12\e{18}$&$\unit{mol}{}\unit{Myr}{-1}$\\
$k_\text{c}$               & Solubility of CO2 in seawater                                     &$10^7$&$\unit{Pa}{}$\\
$L$                 & Total length of subduction zones and mid ocean ridges             &$6\e{4}$&$\unit{km}{}$\\  
$L_\text{w}$               & Latent heat of water                                              &$2469\e{3}$&$\unit{J}{}\unit{kg}{}$\\
$\bar{m}_\text{cc}$      & Mean molar mass of elements participating in weathering           &$32$&$\unit{g}{}\unit{mol}{-1}$\\
$m_{\text{CO}_2}$   & Molar mas of CO$_2$                                               &$44$&$\unit{g}{}\unit{mol}{-1}$\\
$m_\text{w}$             & Molar mass of water                                               &$18$&$\unit{g}{}\unit{mol}{-1}$\\
$M_{\H_2\O}$       & Moles of water per ocean mass                                      &$7.6\e{22}$&$\unit{mol}{}$\\
$P_{\text{CO}_2\oplus}$ & Present-day Earth partial CO$_2$ pressure                         &$33$&$\unit{Pa}{}$\\
$P_{\text{sat0}}$  & Reference saturation vapor pressure                                &$610$&$\unit{Pa}{}$\\
$P_{\text{sat}\oplus}$  & Present-day Earth saturation vapor pressure                       &$1391$&$\unit{Pa}{}$\\
$R_g$                 & Gas constant                                                      &$8.314$&$\unit{J}{}\unit{K}{-1}\unit{mol}{-1}$\\
$\rho_\text{r}$            & Surface regolith density                                          &$2500$&$\unit{kg}{}\unit{m}{3}$\\
$S_\odot$           & Solar constant                                                    &$1360$&$\unit{W}{}\unit{m}{2}$\\
$\sigma$            & Stefan-Boltzmann constant                                         &$5.67\e{-8}$&$\unit{W}{}\unit{m}{-2}\unit{K}{-4}$\\
$t_\odot$           & Age of the Sun                                                    & $4.5$ & Gyr\\
$T_{\text{e}\oplus}$           & Present-day Earth effective temperature                           &$254$&$\unit{K}{}$\\
$T_{\text{s}\oplus}$           & Present-day Earth average surface temperature                     &$285$&$\unit{K}{}$\\
$T_\text{sat0}$     & Saturation pressure reference temperature                         &$273$&$\unit{K}{}$\\
$v_{\text{p}\oplus}$           & Present-day Earth average plate speed                             &$5$&$\unit{cm}{}\unit{yr}{-1}$\\
$\chi_\text{cc}$         & Average fraction of cations participating in weathering           &$0.08$&-\\
\hline
\captionsetup{width=.9\textwidth}
    \caption{Input parameters used for the carbon cycle model presented in Sect. \ref{sec:2.2} and Appendix \ref{sec:AB}. Parameters values are taken from \cite{Foley2015} and Foley, private communication ($a_1$ and $a_2$ parameters).}
    \label{tab:CCTable}
    \end{longtable}
\twocolumn

\end{appendix}

\end{document}